\documentclass{aastex631}

\usepackage{threeparttable}
\usepackage{color}
\usepackage{comment}
\usepackage{amsmath}
\usepackage[normalem]{ulem}

\graphicspath{{./}{figures/}}
\usepackage{subfiles}
\usepackage{subfigure}

\submitjournal{ApJ Letters}

\shorttitle{Neutrino Emission from Seyfert Galaxies}
\shortauthors{IceCube Collaboration}

\begin{document}

\email{analysis@icecube.wisc.edu}

\title{IceCube Search for Neutrino Emission from X-ray Bright Seyfert Galaxies}

\affiliation{III. Physikalisches Institut, RWTH Aachen University, D-52056 Aachen, Germany}
\affiliation{Department of Physics, University of Adelaide, Adelaide, 5005, Australia}
\affiliation{Dept. of Physics and Astronomy, University of Alaska Anchorage, 3211 Providence Dr., Anchorage, AK 99508, USA}
\affiliation{Dept. of Physics, University of Texas at Arlington, 502 Yates St., Science Hall Rm 108, Box 19059, Arlington, TX 76019, USA}
\affiliation{CTSPS, Clark-Atlanta University, Atlanta, GA 30314, USA}
\affiliation{School of Physics and Center for Relativistic Astrophysics, Georgia Institute of Technology, Atlanta, GA 30332, USA}
\affiliation{Dept. of Physics, Southern University, Baton Rouge, LA 70813, USA}
\affiliation{Dept. of Physics, University of California, Berkeley, CA 94720, USA}
\affiliation{Lawrence Berkeley National Laboratory, Berkeley, CA 94720, USA}
\affiliation{Institut f{\"u}r Physik, Humboldt-Universit{\"a}t zu Berlin, D-12489 Berlin, Germany}
\affiliation{Fakult{\"a}t f{\"u}r Physik {\&} Astronomie, Ruhr-Universit{\"a}t Bochum, D-44780 Bochum, Germany}
\affiliation{Universit{\'e} Libre de Bruxelles, Science Faculty CP230, B-1050 Brussels, Belgium}
\affiliation{Vrije Universiteit Brussel (VUB), Dienst ELEM, B-1050 Brussels, Belgium}
\affiliation{Department of Physics and Laboratory for Particle Physics and Cosmology, Harvard University, Cambridge, MA 02138, USA}
\affiliation{Dept. of Physics, Massachusetts Institute of Technology, Cambridge, MA 02139, USA}
\affiliation{Dept. of Physics and The International Center for Hadron Astrophysics, Chiba University, Chiba 263-8522, Japan}
\affiliation{Department of Physics, Loyola University Chicago, Chicago, IL 60660, USA}
\affiliation{Dept. of Physics and Astronomy, University of Canterbury, Private Bag 4800, Christchurch, New Zealand}
\affiliation{Dept. of Physics, University of Maryland, College Park, MD 20742, USA}
\affiliation{Dept. of Astronomy, Ohio State University, Columbus, OH 43210, USA}
\affiliation{Dept. of Physics and Center for Cosmology and Astro-Particle Physics, Ohio State University, Columbus, OH 43210, USA}
\affiliation{Niels Bohr Institute, University of Copenhagen, DK-2100 Copenhagen, Denmark}
\affiliation{Dept. of Physics, TU Dortmund University, D-44221 Dortmund, Germany}
\affiliation{Dept. of Physics and Astronomy, Michigan State University, East Lansing, MI 48824, USA}
\affiliation{Dept. of Physics, University of Alberta, Edmonton, Alberta, T6G 2E1, Canada}
\affiliation{Erlangen Centre for Astroparticle Physics, Friedrich-Alexander-Universit{\"a}t Erlangen-N{\"u}rnberg, D-91058 Erlangen, Germany}
\affiliation{Physik-department, Technische Universit{\"a}t M{\"u}nchen, D-85748 Garching, Germany}
\affiliation{D{\'e}partement de physique nucl{\'e}aire et corpusculaire, Universit{\'e} de Gen{\`e}ve, CH-1211 Gen{\`e}ve, Switzerland}
\affiliation{Dept. of Physics and Astronomy, University of Gent, B-9000 Gent, Belgium}
\affiliation{Dept. of Physics and Astronomy, University of California, Irvine, CA 92697, USA}
\affiliation{Karlsruhe Institute of Technology, Institute for Astroparticle Physics, D-76021 Karlsruhe, Germany}
\affiliation{Karlsruhe Institute of Technology, Institute of Experimental Particle Physics, D-76021 Karlsruhe, Germany}
\affiliation{Dept. of Physics, Engineering Physics, and Astronomy, Queen's University, Kingston, ON K7L 3N6, Canada}
\affiliation{Department of Physics {\&} Astronomy, University of Nevada, Las Vegas, NV 89154, USA}
\affiliation{Nevada Center for Astrophysics, University of Nevada, Las Vegas, NV 89154, USA}
\affiliation{Dept. of Physics and Astronomy, University of Kansas, Lawrence, KS 66045, USA}
\affiliation{Centre for Cosmology, Particle Physics and Phenomenology - CP3, Universit{\'e} catholique de Louvain, Louvain-la-Neuve, Belgium}
\affiliation{Department of Physics, Mercer University, Macon, GA 31207-0001, USA}
\affiliation{Dept. of Astronomy, University of Wisconsin{\textemdash}Madison, Madison, WI 53706, USA}
\affiliation{Dept. of Physics and Wisconsin IceCube Particle Astrophysics Center, University of Wisconsin{\textemdash}Madison, Madison, WI 53706, USA}
\affiliation{Institute of Physics, University of Mainz, Staudinger Weg 7, D-55099 Mainz, Germany}
\affiliation{Department of Physics, Marquette University, Milwaukee, WI 53201, USA}
\affiliation{Institut f{\"u}r Kernphysik, Westf{\"a}lische Wilhelms-Universit{\"a}t M{\"u}nster, D-48149 M{\"u}nster, Germany}
\affiliation{Bartol Research Institute and Dept. of Physics and Astronomy, University of Delaware, Newark, DE 19716, USA}
\affiliation{Dept. of Physics, Yale University, New Haven, CT 06520, USA}
\affiliation{Columbia Astrophysics and Nevis Laboratories, Columbia University, New York, NY 10027, USA}
\affiliation{Dept. of Physics, University of Oxford, Parks Road, Oxford OX1 3PU, United Kingdom}
\affiliation{Dipartimento di Fisica e Astronomia Galileo Galilei, Universit{\`a} Degli Studi di Padova, I-35122 Padova PD, Italy}
\affiliation{Dept. of Physics, Drexel University, 3141 Chestnut Street, Philadelphia, PA 19104, USA}
\affiliation{Physics Department, South Dakota School of Mines and Technology, Rapid City, SD 57701, USA}
\affiliation{Dept. of Physics, University of Wisconsin, River Falls, WI 54022, USA}
\affiliation{Dept. of Physics and Astronomy, University of Rochester, Rochester, NY 14627, USA}
\affiliation{Department of Physics and Astronomy, University of Utah, Salt Lake City, UT 84112, USA}
\affiliation{Dept. of Physics, Chung-Ang University, Seoul 06974, Republic of Korea}
\affiliation{Oskar Klein Centre and Dept. of Physics, Stockholm University, SE-10691 Stockholm, Sweden}
\affiliation{Dept. of Physics and Astronomy, Stony Brook University, Stony Brook, NY 11794-3800, USA}
\affiliation{Dept. of Physics, Sungkyunkwan University, Suwon 16419, Republic of Korea}
\affiliation{Institute of Basic Science, Sungkyunkwan University, Suwon 16419, Republic of Korea}
\affiliation{Institute of Physics, Academia Sinica, Taipei, 11529, Taiwan}
\affiliation{Dept. of Physics and Astronomy, University of Alabama, Tuscaloosa, AL 35487, USA}
\affiliation{Dept. of Astronomy and Astrophysics, Pennsylvania State University, University Park, PA 16802, USA}
\affiliation{Dept. of Physics, Pennsylvania State University, University Park, PA 16802, USA}
\affiliation{Dept. of Physics and Astronomy, Uppsala University, Box 516, SE-75120 Uppsala, Sweden}
\affiliation{Dept. of Physics, University of Wuppertal, D-42119 Wuppertal, Germany}
\affiliation{Deutsches Elektronen-Synchrotron DESY, Platanenallee 6, D-15738 Zeuthen, Germany}

\author[0000-0001-6141-4205]{R. Abbasi}
\affiliation{Department of Physics, Loyola University Chicago, Chicago, IL 60660, USA}

\author[0000-0001-8952-588X]{M. Ackermann}
\affiliation{Deutsches Elektronen-Synchrotron DESY, Platanenallee 6, D-15738 Zeuthen, Germany}

\author{J. Adams}
\affiliation{Dept. of Physics and Astronomy, University of Canterbury, Private Bag 4800, Christchurch, New Zealand}

\author[0000-0002-9714-8866]{S. K. Agarwalla}
\altaffiliation{also at Institute of Physics, Sachivalaya Marg, Sainik School Post, Bhubaneswar 751005, India}
\affiliation{Dept. of Physics and Wisconsin IceCube Particle Astrophysics Center, University of Wisconsin{\textemdash}Madison, Madison, WI 53706, USA}

\author[0000-0003-2252-9514]{J. A. Aguilar}
\affiliation{Universit{\'e} Libre de Bruxelles, Science Faculty CP230, B-1050 Brussels, Belgium}

\author[0000-0003-0709-5631]{M. Ahlers}
\affiliation{Niels Bohr Institute, University of Copenhagen, DK-2100 Copenhagen, Denmark}

\author[0000-0002-9534-9189]{J.M. Alameddine}
\affiliation{Dept. of Physics, TU Dortmund University, D-44221 Dortmund, Germany}

\author{N. M. Amin}
\affiliation{Bartol Research Institute and Dept. of Physics and Astronomy, University of Delaware, Newark, DE 19716, USA}

\author[0000-0001-9394-0007]{K. Andeen}
\affiliation{Department of Physics, Marquette University, Milwaukee, WI 53201, USA}

\author[0000-0003-4186-4182]{C. Arg{\"u}elles}
\affiliation{Department of Physics and Laboratory for Particle Physics and Cosmology, Harvard University, Cambridge, MA 02138, USA}

% ...

\begin{abstract}
The recent IceCube detection of TeV neutrino emission from the nearby active galaxy NGC  1068 suggests that active galactic nuclei (AGN) could make a sizable contribution to the diffuse flux of astrophysical neutrinos. The absence of TeV $\gamma$-rays from NGC 1068 indicates neutrino production in the vicinity of the supermassive black hole, where the high radiation density leads to $\gamma$-ray attenuation. Therefore, any potential neutrino emission from similar sources is not expected to correlate with high-energy $\gamma$-rays. Disk-corona models predict neutrino emission from Seyfert galaxies to correlate with keV X-rays, as they are tracers of coronal activity. Using through-going track events from the Northern Sky recorded by IceCube between 2011 and 2021, we report results from a search for individual and aggregated neutrino signals from 27 additional Seyfert galaxies that are contained in the BAT AGN Spectroscopic Survey (BASS). Besides the generic single power-law, we evaluate the spectra predicted by the disk-corona model. Assuming all sources to be intrinsically similar to NGC  1068, our findings constrain the collective neutrino emission from X-ray bright Seyfert galaxies in the Northern Hemisphere, but, at the same time, show excesses of neutrinos that could be associated with the objects NGC 4151 and CGCG 420-015. These excesses result in a 2.7$\sigma$ significance with respect to background expectations.
\end{abstract}


\section{Introduction} \label{sec:intro}

The IceCube discovery of high-energy cosmic neutrinos \citep{Aartsen:2013bka, Aartsen:2013jdh} demonstrated the feasibility of multimessenger astrophysics to find the origin of cosmic rays (CRs) \citep{Ahlers:2017wkk}. Today, the continuous observation of the high-energy sky by IceCube has revealed evidence for particle acceleration in a nearby Seyfert galaxy, NGC 1068 \citep{IceCube:2022der}. This evidence reinforces the idea that active galactic nuclei (AGN)
can generate very-high-energy CRs \citep{Halzen:1997hw} and are potentially the primary contributors to the diffuse neutrino flux observed by IceCube.
However, the whereabouts of the remaining sources of the high-energy cosmic neutrino flux remain unknown.

NGC 1068 was identified as the most significant source in the analysis of nine years of IceCube neutrino observations in the Northern Hemisphere \citep{IceCube:2022der}. The search identified an excess of 79 events in the direction of NGC 1068 corresponding to a muon neutrino flux of $5\times 10^{-14} \, \rm GeV^{-1}\, cm^{-2}\, s^{-1}$ at 1 TeV with the best-fit power-law spectral index of $3.2$. The neutrino flux measured from NGC 1068 is much larger than the $\gamma$-ray emission in TeV energies if extrapolated from the $\sim$GeV $\gamma$-ray emission measured by {\em Fermi} Large Area Telescope (LAT)~\citep{Fermi-LAT:2019yla,Ballet:2020hze}. 
It is also more than an order of magnitude larger than the upper limits placed by MAGIC and HAWC~\citep{MAGIC:2019fvw, willox2022hawc} on $\sim$ TeV $\gamma$-ray emissions.  High-energy neutrinos and $\gamma$-rays are simultaneously produced whenever CRs interact with ambient matter or radiation within or near cosmic accelerators~\citep{Halzen:2019qkf}.  The observed difference between neutrinos and $\gamma$-rays from NGC 1068 can not be explained by absorption by the extragalactic background light (EBL)~\citep{Murase:2022dog}. Therefore, the environments where the neutrinos are produced in NGC 1068 must be opaque to GeV--TeV $\gamma$-rays that would otherwise accompany the neutrinos. Among the primary candidates are the cores of AGN, which can simultaneously accommodate the efficient production of high-energy neutrinos and offer an optically thick zone that obscures the accompanying $\gamma$-rays. Consequently, the observed $\sim$ GeV $\gamma$-rays should have a different origin than the observed neutrinos~\citep{Murase:2019vdl}.

The high level of neutrino emission compared to $\gamma$-rays from NGC 1068 is in agreement with the multimessenger picture for the total diffuse high-energy neutrino flux reported by IceCube \citep{IceCube:2021uhz, Aartsen:2020aqd}. This isotropic neutrino flux is an order of magnitude greater at medium energies ($\sim 30$ TeV) than it is at very high energies ($> 100$ TeV) \citep{Aartsen:2020aqd}. 
Comparisons between the diffuse neutrino flux at medium energies and the isotropic gamma-ray background observed by {\em Fermi}~\citep{Ackermann:2014usa} independently suggest neutrino production in environments that are obscured to high-energy $\gamma$ rays ~\cite{Senno:2015tra, Murase:2015xka, Bechtol:2015uqb, Capanema:2020rjj, Fang:2022trf}, in line with the previously discussed multi-messenger picture of NGC 1068.

AGN host supermassive black holes at their centers that power these galaxies via the release of gravitational energy from accreting matter. 
The accreting matter falls toward the black hole and forms a disk around the central engine, i.e., the core of AGN. The cores of AGN are optically thick for GeV--TeV $\gamma$-rays. Simultaneously, they provide the target matter and radiation that is required for the efficient production of neutrinos. See \citep{Murase:2022feu} and references therein for more details.

IceCube has previously searched for the collective neutrino flux from AGN cores by considering a large catalog of AGN and assuming that neutrinos are produced by accelerated cosmic rays in the AGN accretion disks. \citep{IceCube:2021pgw}. In the study presented here, motivated by the observation of neutrino emission from NGC 1068, we instead aim to identify neutrino emission from a targeted selection of X-ray bright Seyfert galaxies in the Northern hemisphere (Dec $>-5^\circ$). Here, we turn our attention to neutrino emission from the coronae of Seyfert galaxies \citep{Murase:2019vdl, Kheirandish:2021wkm} and select bright sources based on the intrinsic X-ray flux (2--10 keV) to identify sources similar to NGC 1068. Each candidate source is searched individually for signs of point-like neutrino emission above background expectations.
We employ model-predicted neutrino spectra in addition to the generic single power-law in this search. Additionally, we perform a joint analysis of all sources using a {\em stacking} method to search for potential collective emissions from these sources. 

In the next section, we discuss neutrino emission from the coronae of bright Seyfert galaxies. In Section \ref{sec:method} we present details about source selection and analysis method. Results of each test performed in this study are presented in Section \ref{sec:results}. In Section \ref{sec:discussion} we discuss the implication of each analysis.

\section{Seyfert galaxies as sources of high-energy neutrinos} \label{sec:nuseyferts}
In Seyfert galaxies, accretion dynamics and magnetic dissipation lead to the formation of a hot, highly magnetized, and turbulent corona above the disk, see e.g.~\cite{Miller:1999ix}. The dense environments near the supermassive black holes together with the acceleration of CRs in the coronae offer suitable conditions for the production of high-energy neutrinos. Simultaneously, these systems will be opaque to the accompanying very-high-energy $\gamma$-rays. Such a scenario has been examined by models that attempt to describe the neutrino flux at the medium energy range of the IceCube high-energy diffuse astrophysical neutrino observation and the soft best-fitted spectrum reported from NGC 1068~\citep{Murase:2019vdl, Inoue:2019yfs, Kheirandish:2021wkm, Eichmann:2022lxh}. These models, commonly referred to as disk-corona models, can accommodate the high level of neutrino emission at medium energies and the measured neutrino flux from NGC 1068, see e.g.~\cite{Murase:2022feu} for detailed discussion. Here, we focus on the disk-corona model presented by~\cite{Murase:2019vdl, Kheirandish:2021wkm} where neutrino emission is a product of stochastic acceleration in the corona and interaction of CRs with gas or the radiation from the innermost regions of the AGN. On the basis of the reported intrinsic X-ray flux, this model finds NGC 1068 as the brightest source in IceCube and suggests that additional sources might be identified in IceCube if they pose similar characteristics to NGC 1068. 

AGN coronae are primarily characterized by X-ray emission that is powered thermally. In the scenario discussed in the work, the CR injection fraction is proportional to the dissipation rate in the coronae which is determined by the thermal X-ray luminosity. Naturally, the intrinsic X-ray luminosity becomes the principal parameter in the disk-corona models for estimating the neutrino emission. Additional model parameters include the CR to thermal pressure ratio that summarizes the CR budget and the turbulence strength. Larger values of this ratio result in increased neutrino production. While moderate values can explain the diffuse neutrino flux at medium energies, a higher level of CR pressure is needed to explain the neutrino flux measured in the direction of NGC 1068~\citep{Kheirandish:2021wkm}. This assumption is heavily tied to the measured X-ray flux, and we will return to it in the discussion of the results. For this study, we solely focus on the high CR pressure scenario (i.e., $P_{\rm CR}/P_{\rm th}=50\%$), given that identification of sources with moderate CR pressure requires next generation neutrino telescopes~\citep{Kheirandish:2021wkm}.

\section{Searching for neutrino emission from bright Seyfert galaxies in the Northern Sky} \label{sec:method}

\subsection{Source Catalog}
Our source catalog is based on the BAT AGN Spectroscopic Survey (BASS)~\citep{Ricci:2017dhj} which is an all-sky study of AGN detected in the X-ray band. In our selection, we start from all sources identified as Seyfert galaxies with the 105-month $Swift$-BAT classification~\citep{Oh:2018wzc}. BAT sources in the Northern Sky (Dec $>-5^\circ$) are ranked by their intrinsic X-ray fluxes in the 2--10 keV band. 
Sources with weak intrinsic X-ray fluxes are not expected to produce sizeable neutrino fluxes. Therefore, we select only bright sources for our analysis with intrinsic X-ray fluxes that are at least 10~\% of that of NGC~1068, which is approximately the sensitivity of IceCube in the Northern Sky according to the previous result of NGC 1068~\citep{IceCube:2022der}. 
The selection retains 28 sources including NGC 1068. Considering the {\em a priori} knowledge of a strong flux from this source, we separate out the results for NGC~1068 to avoid bias in the analysis. Therefore, we discuss the exclusion and inclusion of NGC~1068 separately. To be conservative and to take into account the fact that the remaining sources can still give neutrino signals significant enough based on the model, we draw our conclusion without NGC~1068, and the results including NGC~1068 are shown for completeness.    

\subsection{IceCube detector \& data}
The IceCube Neutrino Observatory at the South Pole is a Cherenkov neutrino telescope that utilizes 1 $\rm{km}^3$ volume of glacial ice to detect high-energy neutrinos. The detector is an array of 5,160 digital optical modules (DOMs), each composed of a photomultiplier tube (PMT) and onboard read-out electronics \citep{IceCube:2016zyt}. Cherenkov photons emitted by the relativistic charged particles produced in interactions between neutrinos and nucleons are collected by the PMTs. The photon count and arrival time recorded by each DOM are used to reconstruct the energy and direction of each event. The flavors and interaction types of neutrinos lead to different event morphologies in the detector. Among them, track events resulting from charged-current interactions of muon neutrinos can be  reconstructed with good angular resolutions \citep{IceCube:2022der}, which makes them ideal for pointing back to their sources. However, muons and neutrinos produced in the atmosphere present a significant background. Fortunately, the Earth acts as an effective muon filter for up-going events from the Northern Hemisphere, thus a better sensitivity can be reached due to the suppression of the atmospheric muon background. Here, we analyze the through-going muon tracks in the Northern sky (Dec $>-5^\circ$, \cite{IceCube:2021uhz}).

The data sample is processed in the same way as in~\cite{IceCube:2022der} which has new data processing, data calibration, and event reconstruction implemented that result in substantially improved energy reconstructions and point spread function at low to medium energies. The details can be found in the supplementary materials of~\cite{IceCube:2022der}. We extend the data used in this previous work by adding extra 1.7 years of experimental data. We also improved the modeling of the muon contribution from the decay of tau leptons produced by tau neutrino charged-current interactions outside the detector by adding dedicated Monte Carlo (MC) simulations. The extension of the lifetime increases the total number of events by $\sim$20\%. The experimental data starts with the fully built and commissioned detector (aka IC86) on 
2011-05-13 and ends on 2022-02-13 with a total live time of 3804 days, corresponding to 794,301 track events in total.

\subsection{Analysis}
\begin{figure}[t!]
\centering
\includegraphics[width=0.65\textwidth]{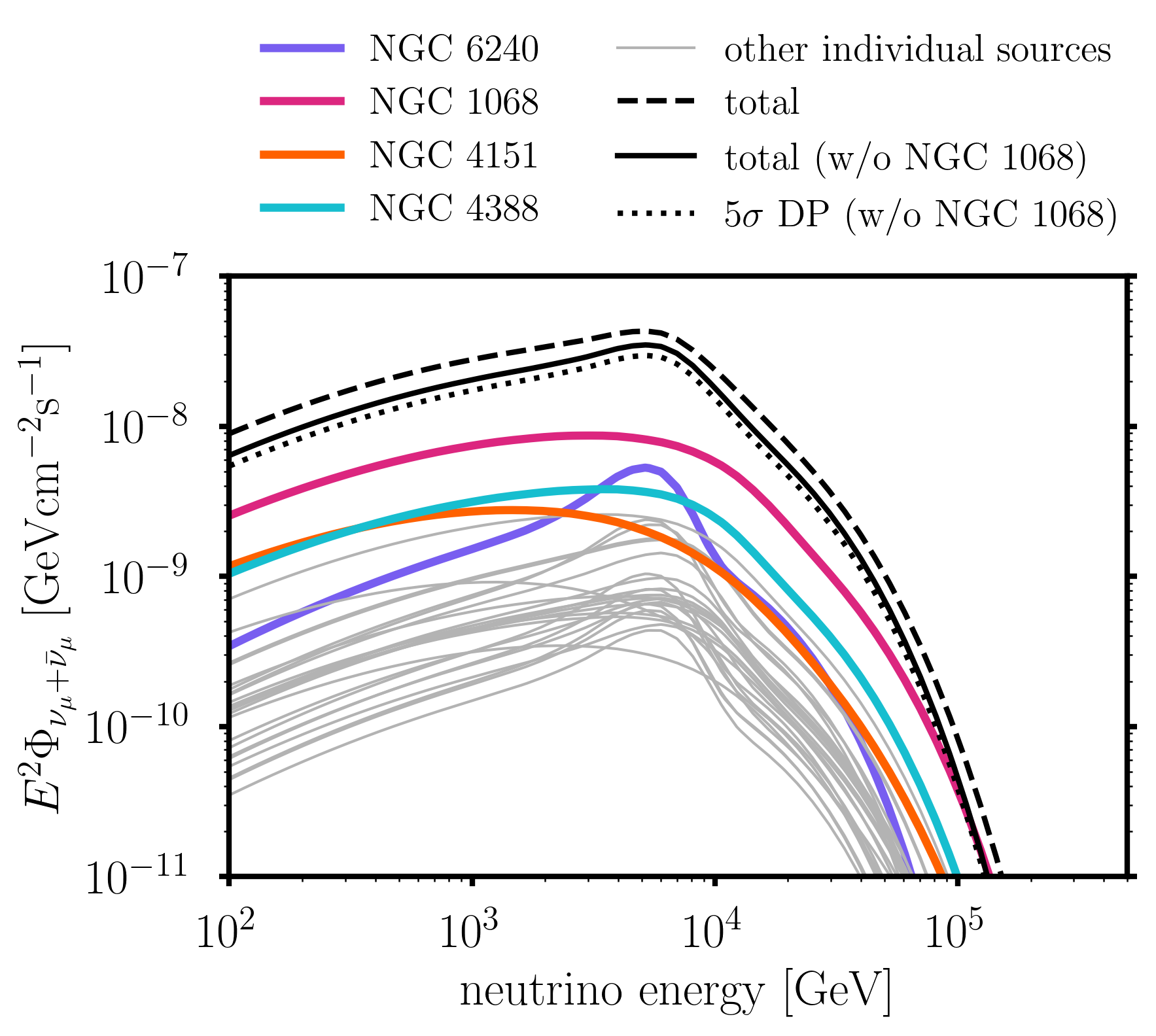}
\caption{The expected flux of each source (thin lines) from the disk-corona model (high CR pressure) with the top 4 sources highlighted. The total fluxes excluding or including NGC 1068 are shown. The $5\sigma$ discovery potential (DP), which excludes NGC 1068, is given as well.}
\label{fig:model_flux_and_sensitivity}
\end{figure}
\begin{deluxetable*}{lDDDDDDDD}[hbt!]
\tablecaption{Results}
\tablewidth{20pt}
\tablehead{
\colhead{} & \multicolumn2c{spectral model}  & \multicolumn2c{$n_{\rm exp}$} &  \multicolumn2c{TS} & \multicolumn2c{$\hat{n}_{\rm s}$} & \multicolumn2c{$\hat{\gamma}$} & \multicolumn2c{$p_{\rm local}$} & \multicolumn2c{$p_{\rm global}$} &\multicolumn2c{$n_{\rm UL}$} 
}
\decimals
\startdata
Stacking Searches & \phantom{$-$} & \phantom{$-$} & \phantom{$-$} & \phantom{$-$} & \phantom{$-$} & \phantom{$-$} & \phantom{$-$} & \phantom{$-$} \\\hline
Stacking (excl.) & \textnormal{disk-corona} & 154.0 & 0.1 & 5 & - & $2.4\times 10^{-1}\,(0.7\,\sigma)$ & $2.4\times 10^{-1}\,(0.7\,\sigma)$ & 51.1 \\
Stacking (incl.) $^{(*)}$ &  \textnormal{disk-corona} & 199.0 & 11.2 & 77 & - & $1.1\times 10^{-4}\,(3.7\,\sigma)$ & $-$ & 128.0 \\
\hline
Catalog Search 1 & \phantom{$-$} & \phantom{$-$} & \phantom{$-$} & \phantom{$-$} & \phantom{$-$} & \phantom{$-$} & \phantom{$-$} & \phantom{$-$} \\\hline
CGCG 420-015 & \textnormal{disk-corona} & 3.2  & 11.0 & 31 & - & $2.4\times 10^{-4}\,(3.5\,\sigma)$ & $6.5\times 10^{-3}\,(2.5\,\sigma)$ & 46.4 \\
NGC  4151 & \textnormal{disk-corona} & 13.1  & 9.0  & 23 & - & $6.4\times 10^{-4}\,(3.2\,\sigma)$ & $-$ & 39.5 \\
NGC  1068 $^{(*)}$ & \textnormal{disk-corona} & 44.6 & 23.4 & 48 & - & $3.0\times 10^{-7}\,(5.0$ $\sigma)$& $-$ & 61.4 \\
\hline
Catalog Search 2 & \phantom{$-$} & \phantom{$-$} & \phantom{$-$} & \phantom{$-$} & \phantom{$-$} & \phantom{$-$} & \phantom{$-$} & \phantom{$-$} \\\hline
NGC  4151  & \textnormal{power-law} & $-$ & 7.4 & 30 & 2.7 & $6.4\times 10^{-4}\,(3.2\,\sigma)$ & $1.7\times 10^{-2}\,(2.1\,\sigma)$ & 61.4 \\
CGCG 420-015 & \textnormal{power-law} & $-$ & 9.2  & 35 & 2.8 & $3.0\times 10^{-3}\,(2.7\,\sigma)$ & $-$ & 62.1 \\
NGC  1068 $^{(*)}$ & \textnormal{power-law} & $-$ & 29.5 & 94 & 3.3 & $8.0\times 10^{-8}\,(5.2\,\sigma)$ & $-$ & 94.9 \\
\enddata
\tablecomments{Results for the stacking search and selected results from two catalog searches, Catalog Search 1: disk-corona model; and Catalog Search 2: power-law model. Best-fitted TS, $\hat{n}_s$, local (pre-trial) and global (post-trial) $p$-values, and corresponding significances are shown. For the disk-corona model analysis, expected numbers of events ($n_{\rm exp}$) are listed and for the power-law analysis, best-fitted spectral indices $\hat{\gamma}$ are listed. $n_{\rm UL}$ column shows the 90\% upper limits of the numbers of signal events. Upper limits assuming power-law spectra are given assuming $E^{-3}$. Results marked with $^{(*)}$ are provided for completeness but are not used to compute final significances because evidence for neutrino emission from NGC 1068 was known prior to this work~\citep{IceCube:2022der, IceCube:2019cia}. 
} 
\label{tab:results}
\end{deluxetable*}
\begin{figure*}
    \centering
     \subfigure{%
        \includegraphics[width=0.315\linewidth]{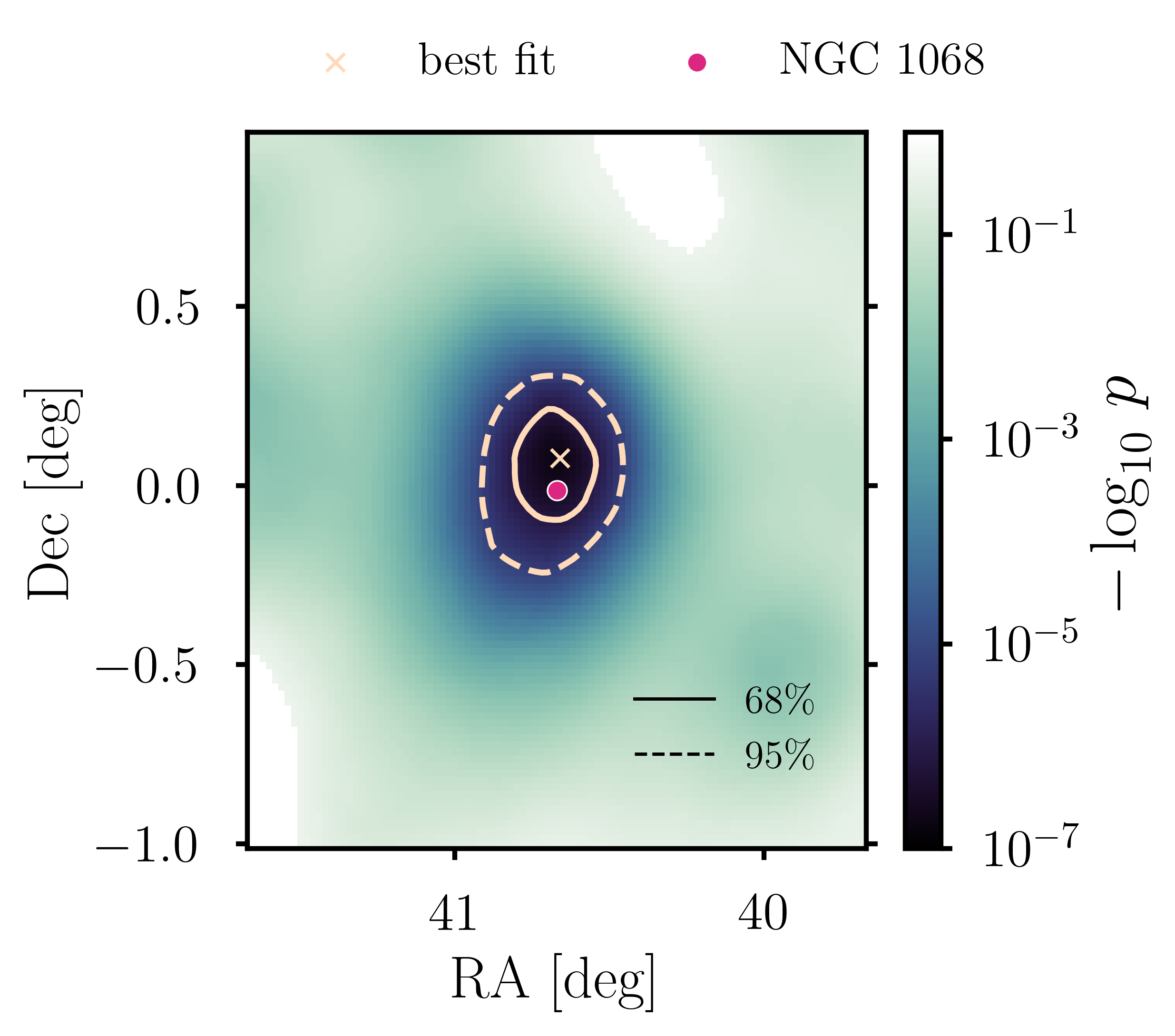}
        \label{fig:gull}}
    \subfigure{
        \includegraphics[width=0.315\linewidth]{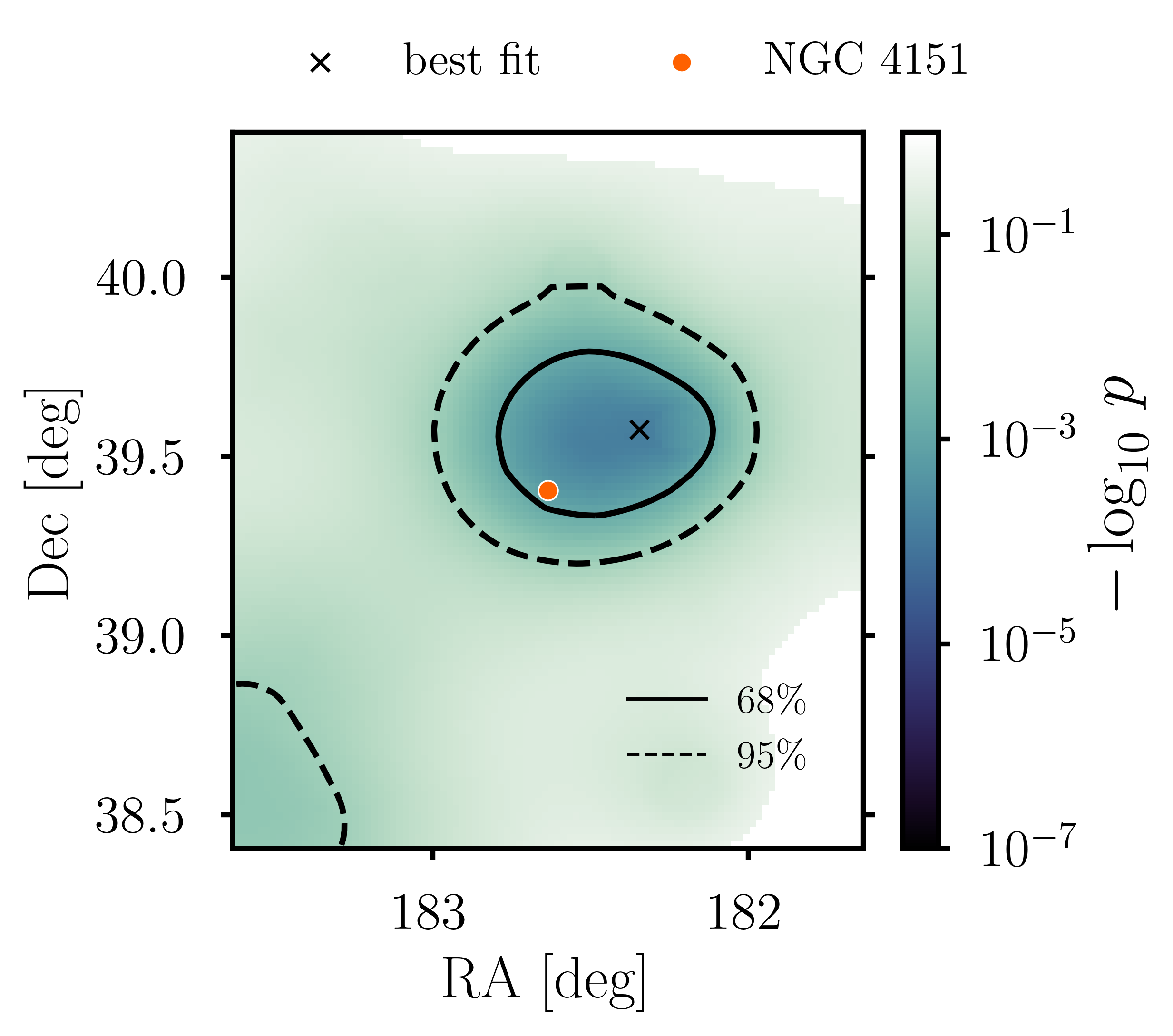}
        \label{fig:tiger}}
    \subfigure{
        \includegraphics[width=0.315\linewidth]{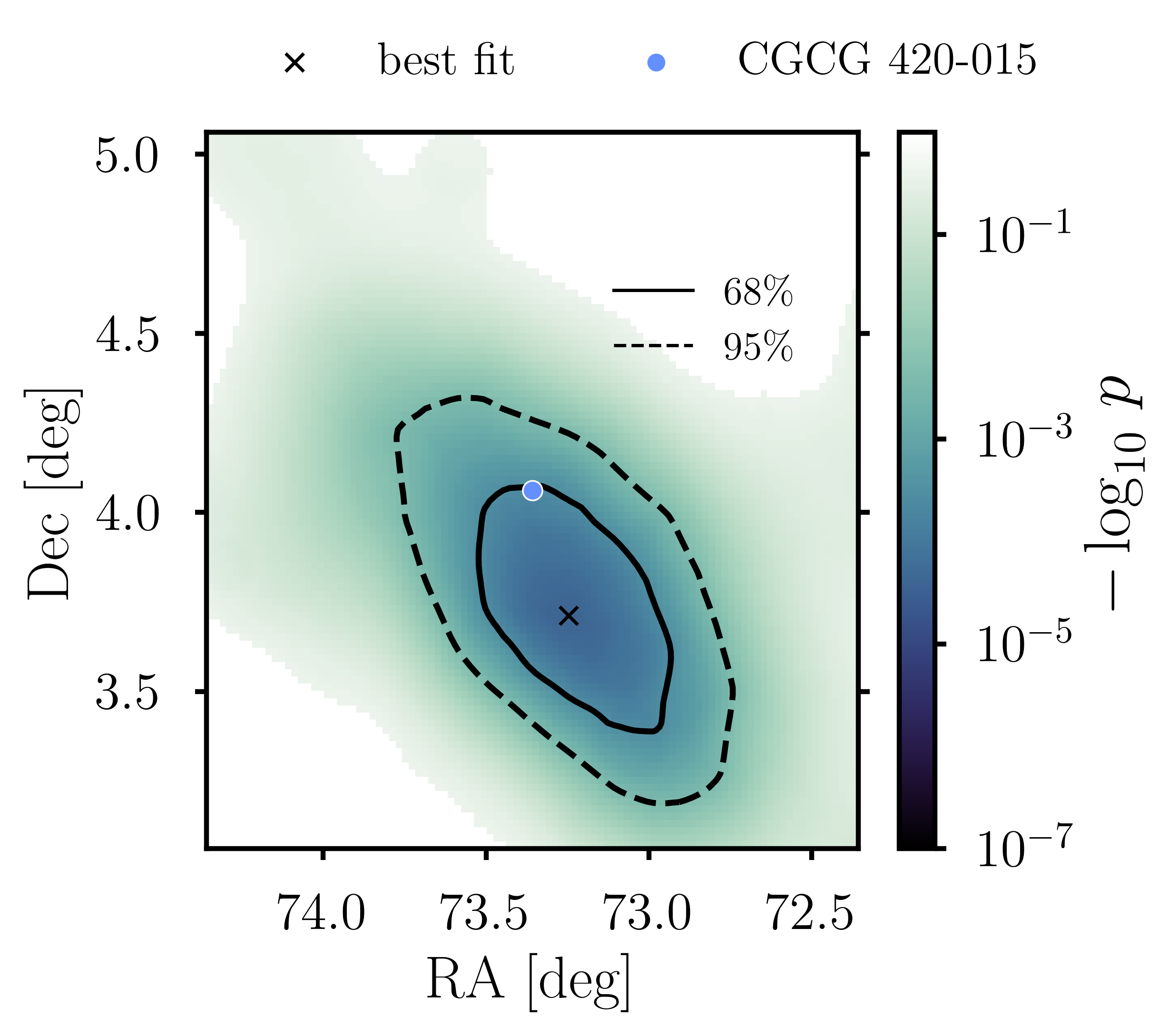}
        \label{fig:tiger}}\\     
    \subfigure{%
        \includegraphics[width=0.315\linewidth]{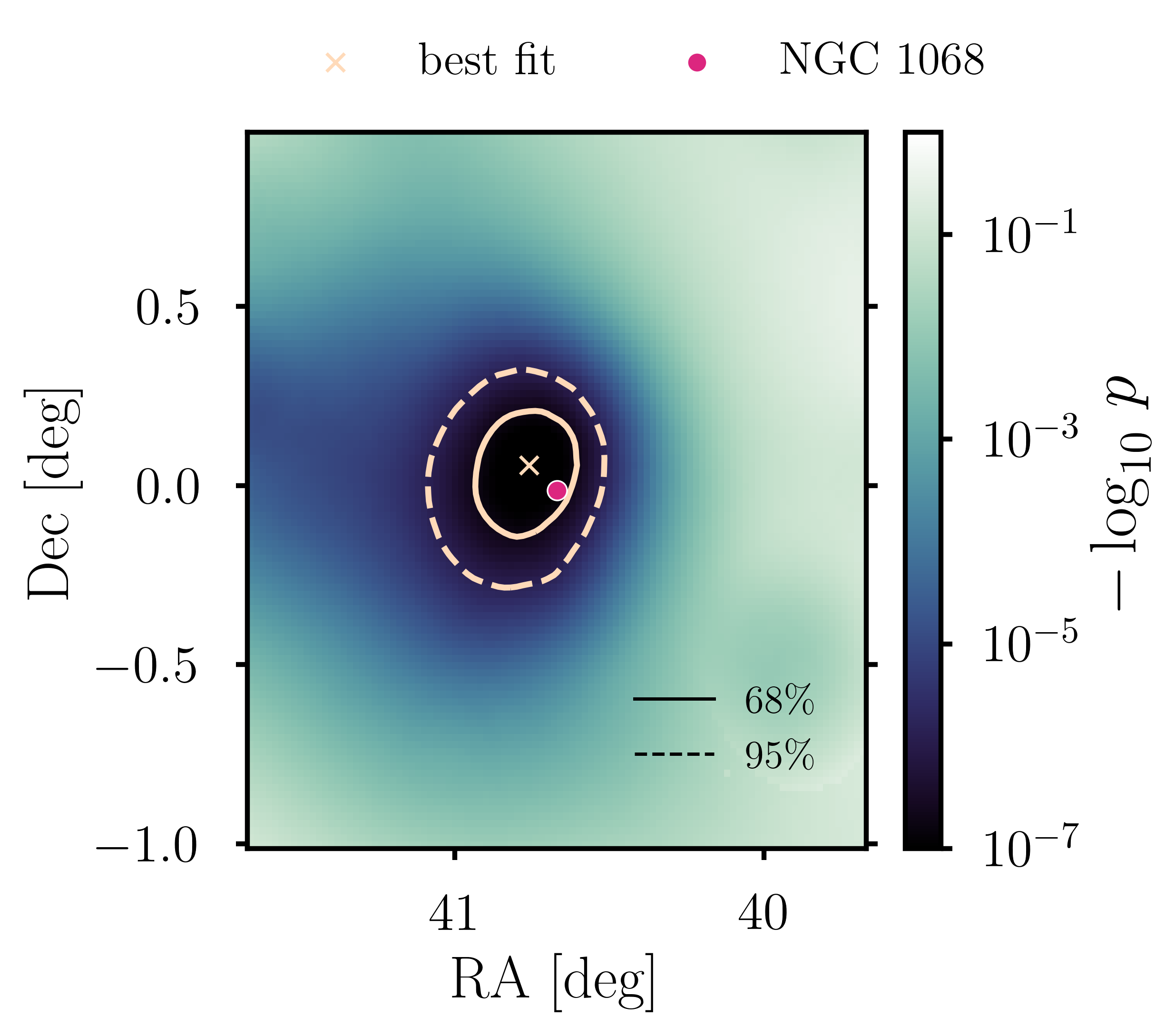}
        \label{fig:gull}}
    \subfigure{
        \includegraphics[width=0.315\linewidth]{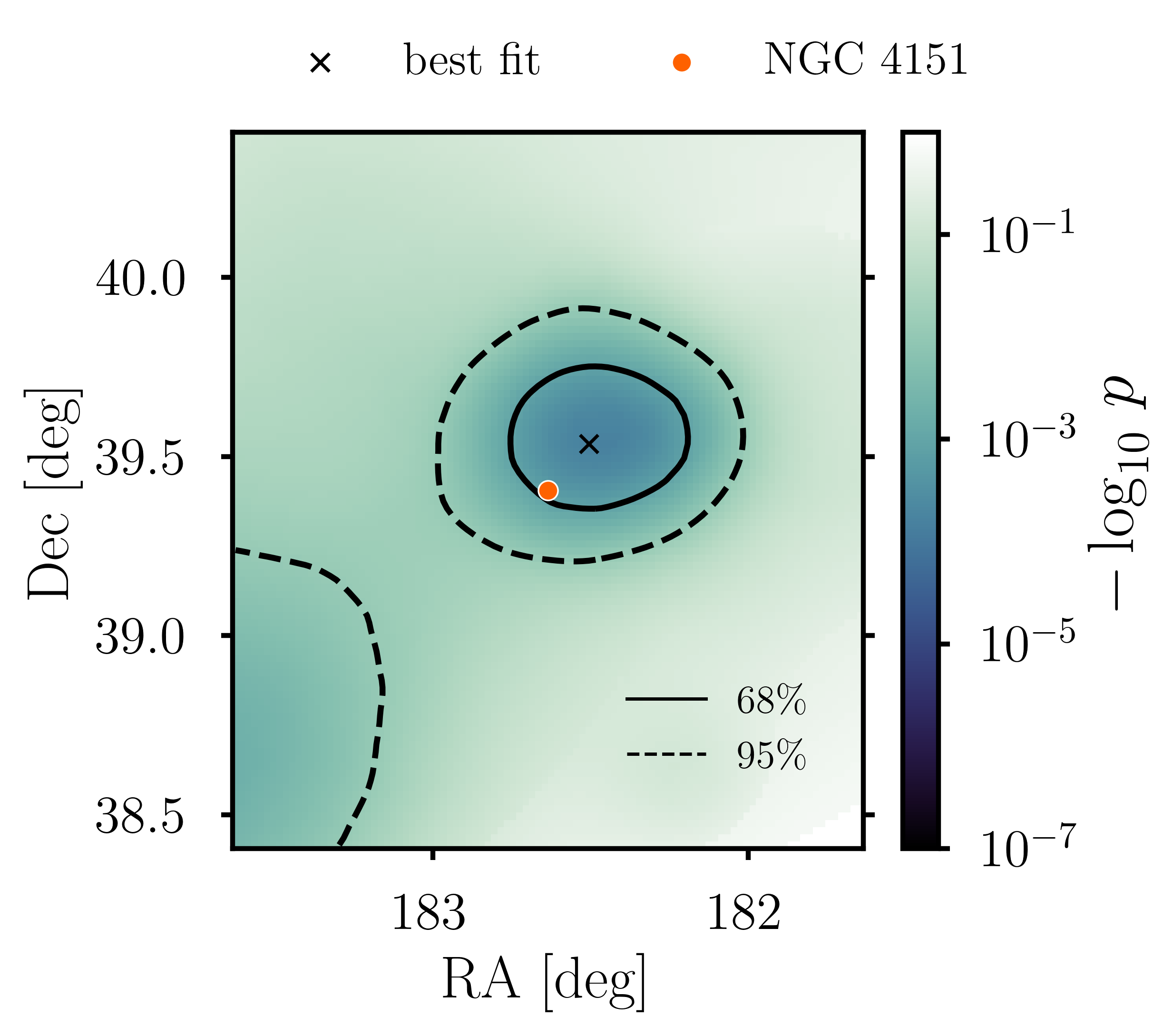}
        \label{fig:tiger}}
    \subfigure{
        \includegraphics[width=0.315\linewidth]{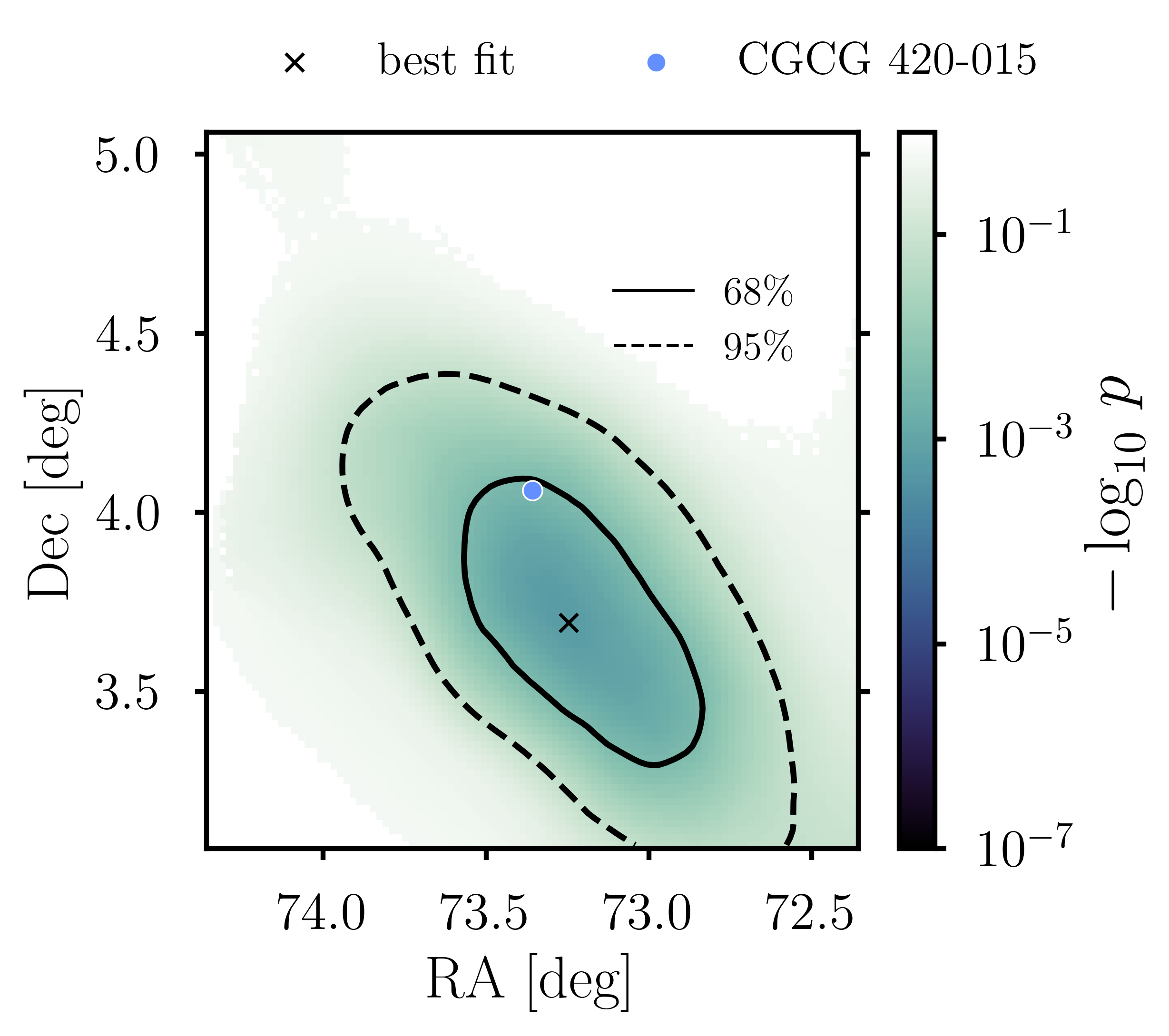}
        \label{fig:tiger}}
    \caption{Local (pre-trial) $p$-value maps around the most significance sources NGC 1068 (left), NGC 4151 (middle), and CGCG 420-015 (right) with the the disk-corona model fit (top) and the power-law fit (bottom). Colored points show the locations of sources and crosses show the best-fit locations. Contours correspond to 68\% (solid) and 95\% (dashed) confidence regions. }
    \label{fig:p_scan}
\end{figure*}
In order to discriminate between potential neutrino emission from our selected sources and the background composed of atmospheric and isotropic astrophysical neutrinos, we employ the unbinned likelihood ratio hypothesis testing method that utilizes the direction, energy proxy, and angular uncertainty of the events~\citep{Braun:2008bg}. We perform two types of searches. One is the catalog search, where we look for the neutrino emission from each source separately, using both, a  power-law assumption and disk-corona model prediction. The outcome will be used to conduct a binomial test to examine the significance of observing excesses for $k$ sources in these two scenarios for our catalog search. This could potentially  identify a sub-set of sources that do individually not pass the significance threshold. The other is the stacking search, where the emission of each selected source is combined according to model expectations in order to obtain an enhanced signal above the background. In the stacking analysis, only the disk-corona model flux is tested. All searches were specified {\em a-priori}.

These analyses apply the improved kernel density estimation (KDE) method  presented in~\cite{IceCube:2022der} to generate probability density functions (PDFs), which improves the modeling of directional distributions of individual neutrinos. These distributions depend on the shape of the spectrum and were therefore generated for each spectral index in the power-law flux analysis~\citep{IceCube:2022der}. Using the same methods, we generate the signal spatial PDFs corresponding to the disk-corona model spectra with an updated KDE generation pipeline, where we utilize a grid in the intrinsic X-ray luminosity $L_X$, the parameter that determines the shape of the flux. See the Appendix for more details about the likelihood method.

For the analyses based on the disk-corona model, we calculate the expected number of events from each source with the high-pressure scenario described in~\cite{Kheirandish:2021wkm}. Here, the flux shape varies with $L_X$, and the flux normalization changes with the CR pressure. Other parameters in the calculation are fixed to values fitting the observed flux from NGC 1068 assuming all sources to be intrinsically similar to NGC 1068. The expected fluxes of selected sources are shown in Fig.~\ref{fig:model_flux_and_sensitivity}. The total model fluxes with and without NGC 1068 for the stacking search are also shown with comparison to the $5\sigma$ discovery potential, where a $6\sigma$ significance is expected even without NGC 1068 for the optimistic scenario. Testing the performance of the analysis shows that if the disk-corona model predicts the true flux, modeling the flux correctly gives a gain of $\sim 1\sigma$ significance for the stacking search and $\sim 0.7\sigma$ for NGC 1068 compared to fitting a power-law spectrum, as can be seen in Fig.~\ref{fig:appendix_fig5} in the Appendix. 

For the catalog search based on the power-law spectrum assumption, we follow  the same procedure as in~\cite{IceCube:2022der}. This analysis is to complement the search discussed above for possible high-energy events if neutrino emission from any of the sources is extended to above $\sim\,$100~TeV, and thus offer an intuitive comparison with other work by applying the generic power-law flux assumption. 

\section{Results} \label{sec:results}

A summary of our results is shown in Table~\ref{tab:results}. In addition to NGC 1068, we find excesses of neutrino emission which could be associated with two other sources: CGCG 420-015 and NGC 4151. CGCG 420-015 is the most significant source in the search based on the disk-corona model with a 3.5$\sigma$ local (pre-trial) significance, while NGC 4151 stands out as the most significant source in the search based on the power-law spectrum assumption with a 3.2$\sigma$ local significance. Starting from the larger value 3.5$\sigma$, the global significance is lowered by the \textit{look-elsewhere effect} by accounting for the number of objects in the catalog (27) and the two spectral assumptions. The global significance of the catalog search is $2.3\sigma$ as determined by repeated applications of the entire analysis to simulated data containing only background events.

\begin{figure*}[t!]
\centering
\subfigure{%
\includegraphics[width=0.32\textwidth]{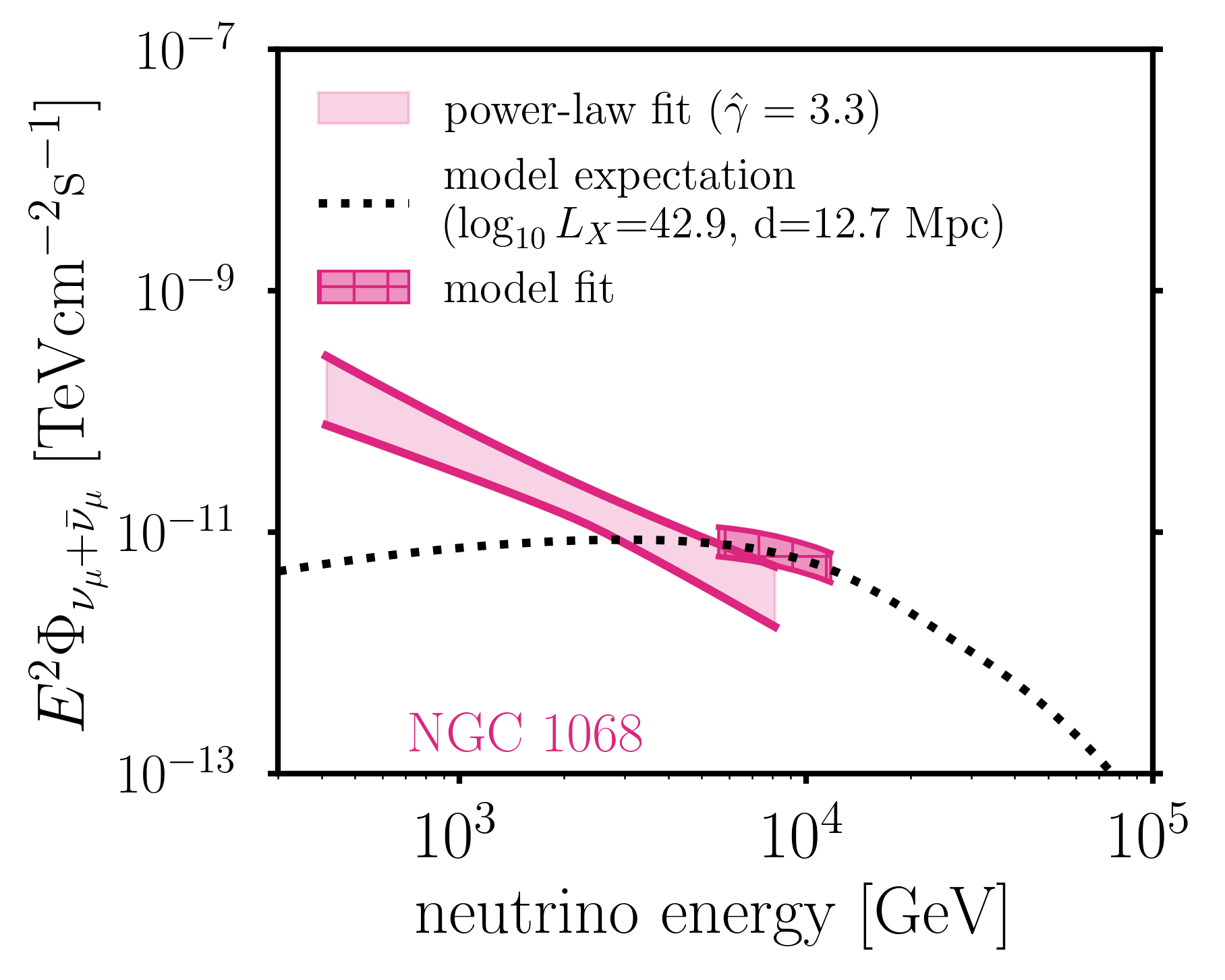}}
\subfigure{%
\includegraphics[width=0.32\textwidth]{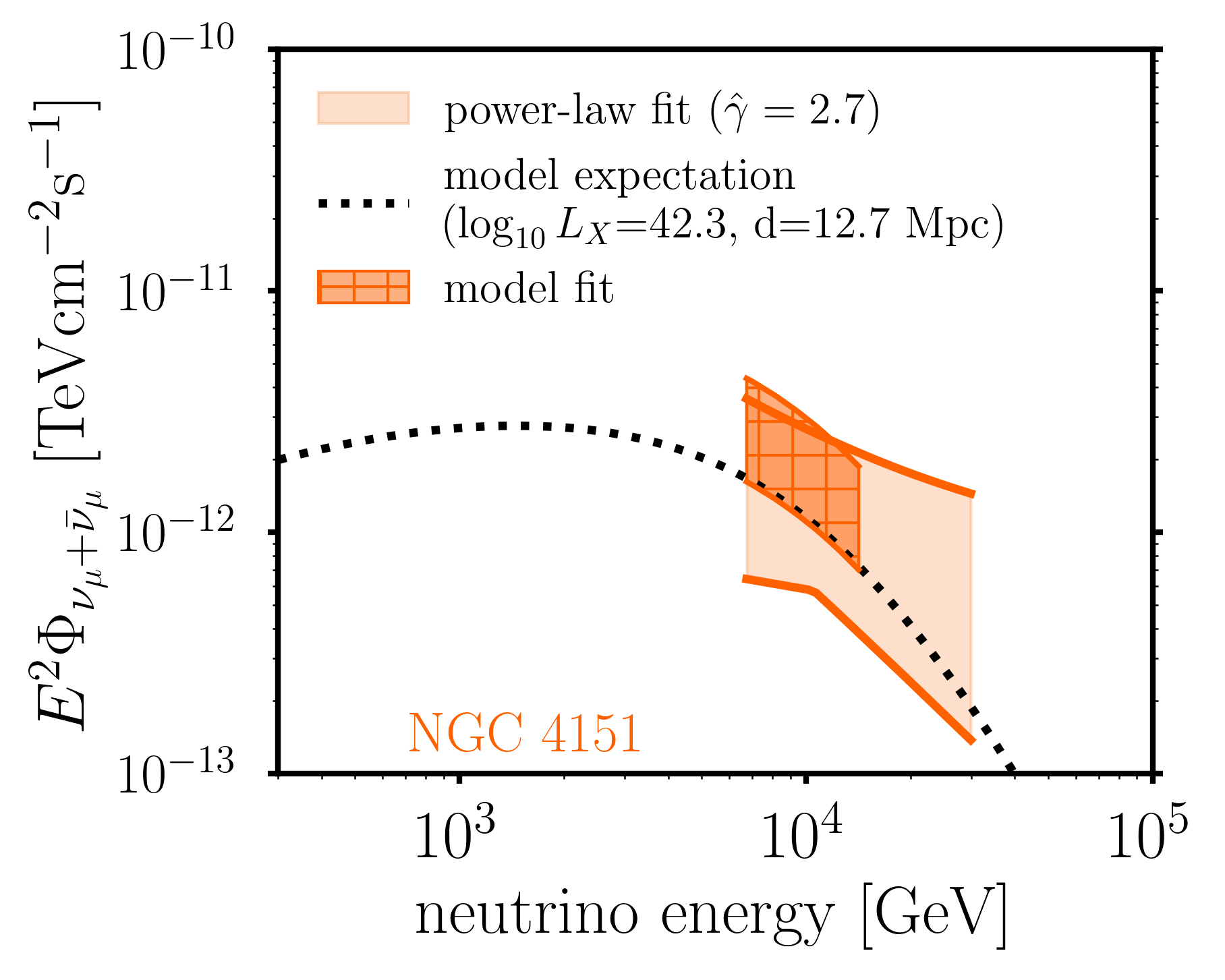}}
\subfigure{%
\includegraphics[width=0.32\textwidth]{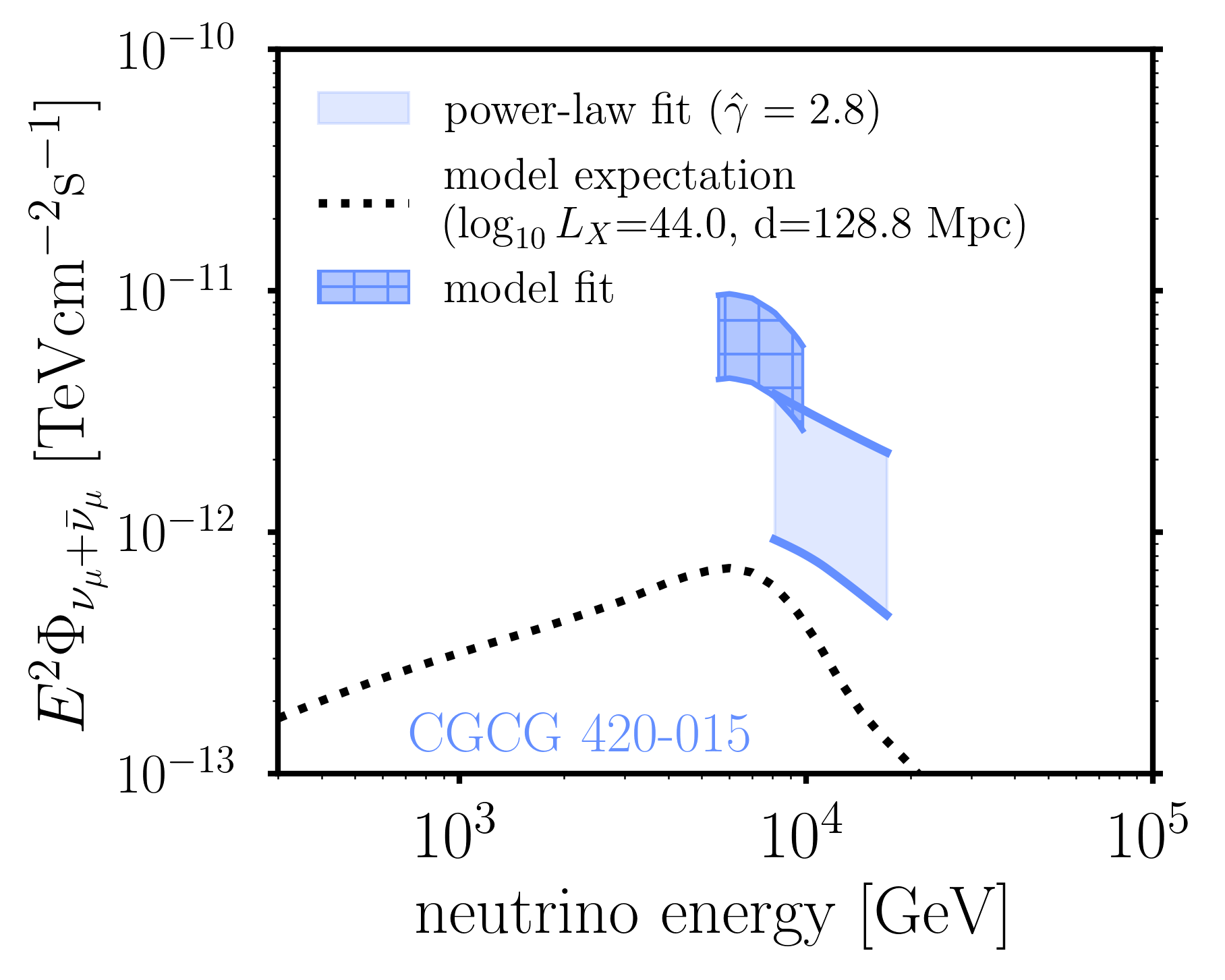}}
\caption{The best-fitted flux and the corresponding 68\% statistical uncertainty for model and power-law searches compared to the expected disk-corona flux for the most significant sources NGC 1068 (left), NGC 4151 (middle), and CGCG 420-015 (right). Distance and intrinsic X-ray luminosity (in erg/s) for each source are taken from BASS. Systematic uncertainties are subdominant. The neutrino energy range of the best-fitted flux is computed from the 68\% central bins contribution to the total TS value. See the Appendix for more details about the determination of the energy range and the description of systematic uncertainties.}
\label{fig:flux_catalog_search}
\end{figure*}

\begin{figure*}[t!]
\centering
\includegraphics[width=0.65\textwidth]{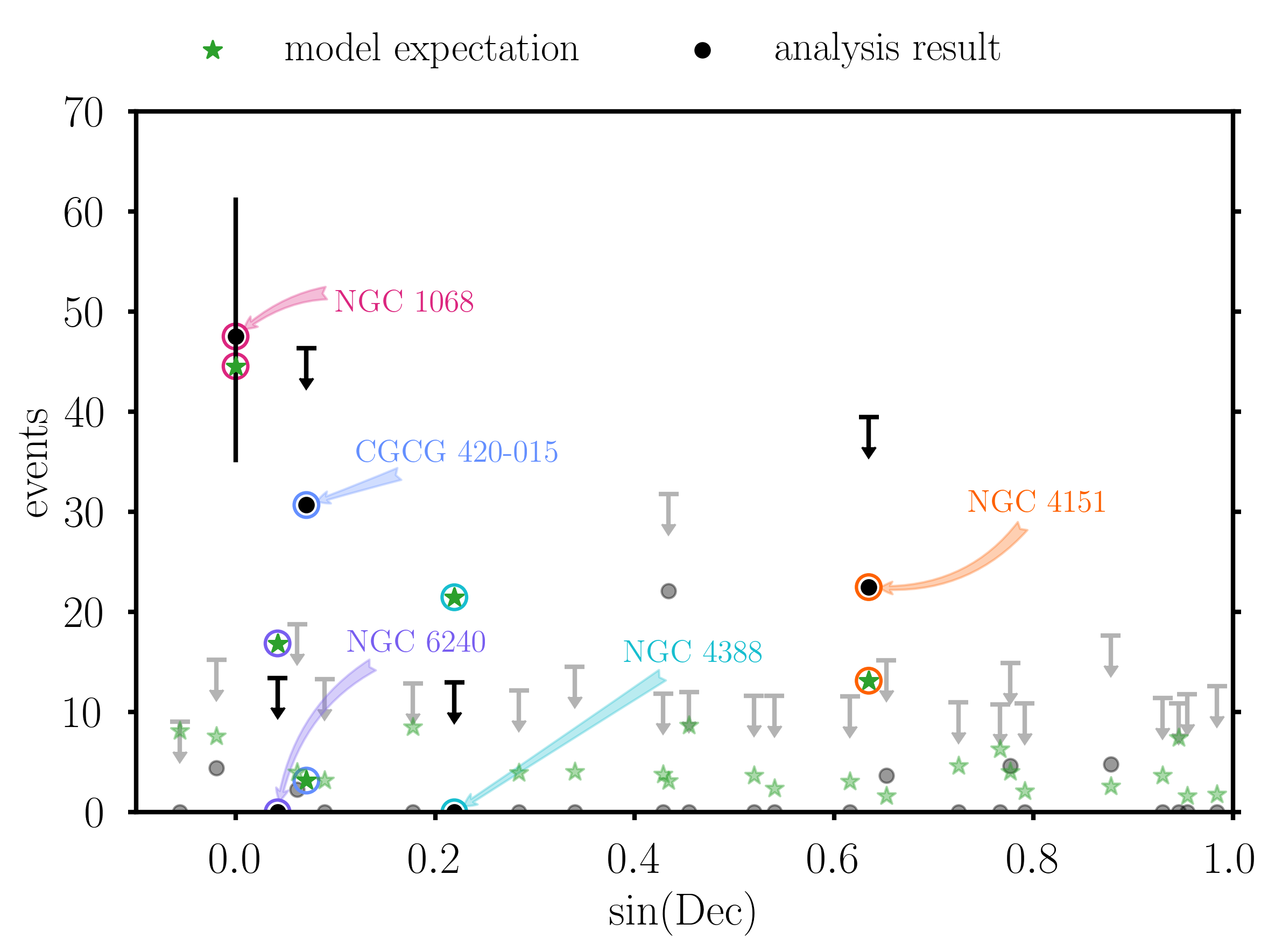}
\caption{Expected numbers of events (green stars) from the model and the best-fitted numbers of signal events (black circles) for individual sources. Down arrows show the 90\% C.L. upper limits. The five candidate sources with the strongest expected neutrino signal (disc-corona model) are highlighted. }
\label{fig:results_summary}
\end{figure*}

The local significance of NGC 1068 increases slightly compared to the previous result in~\cite{IceCube:2022der} due to the extension of the dataset. Assuming the power-law spectral model, we obtain a local $p$-value of $8\times10^{-8}$, which corresponds to a global significance of $4.3\sigma$ in the previous catalog search described in \cite{IceCube:2022der} (N=110 candidate sources) and is consistent with expectations assuming the best-fitting values in the previous analysis. Panels in Fig.~\ref{fig:p_scan} display the $p$-value scan in the nearby region around the most significant sources under our two spectral assumptions. Their best-fit fluxes are shown in Fig.~\ref{fig:flux_catalog_search} where the model fit and power-law fit can be compared. The fluxes given by the model fit can also be compared to their expected spectra. 
For all selected sources, Fig.~\ref{fig:results_summary} presents event numbers of the expectation as well as the measurement with the computed 90\% confidence level upper limits. The full information and results for all sources are tabulated in Table~\ref{tab:int_results}. 

The binomial test \citep{IceCube:2018ndw} allows us to combine these results without any assumption about the relative contribution from the candidate sources in our catalog. Assuming the disk-corona flux model for each candidate source, and excluding NGC 1068, we find a combined pre-trial significance of $2.9\sigma$ in excess of the expected backgrounds. The global significance of this test becomes 2.7$\sigma$, because we tested two different flux models. This result rests on the observation of two candidate sources ($k=2$) with small local $p$-values in our catalog: NGC 4151 and CGCG 420-015. Including NGC 1068 would increase the number of sources, identified in this test, by one ($k=3$), and result in an a-posteriori significance of 4$\sigma$. For more details about the binomial test and this result, see the Appendix.

As can be seen in Table~\ref{tab:results}, there is no significant excess found in the stacking search without contribution from NGC 1068 with $p$-value = 0.24. The $90\%$ C.L. upper limit on the total number of signal neutrinos from the selection of sources excluding NGC 1068 is $n_s^{\rm UL}=51$, while $n_{\rm exp} = 155$ events were expected had all analysis assumptions been met exactly. If NGC 1068 is included, the upper limit becomes $n_s^{\rm UL}=128$ compared to an expectation of $n_{\rm exp}=199$.

\section{Discussion}\label{sec:discussion}
In this study, we probed neutrino emission in the direction of the brightest Seyfert galaxies identified in the BASS catalog. Based on the disk-corona model prediction, collective emission from these sources should emerge with high significance in IceCube data provided that the sources are characterized by a high CR pressure, similar to NGC 1068. On one hand, taking the intrinsic X-ray luminosities reported by BASS ~\citep{Ricci:2017dhj} at face value, the absence of a strong signal in the stacking search implies that the model parameters that are suited to explain the observed flux from NGC 1068 appear not to be shared with most sources in the catalog. On the other hand, the results of the catalog searches, in particular the $2.7\sigma$ neutrino excess in the binomial test could demonstrate not only the existence of a subset of Seyfert galaxies being similar to NGC 1068 which adds to the growing indications that at least a subset of AGN contribute to the high-energy neutrino flux, but also the feasibility of identifying them while more data is needed for a robust identification of more sources. 

The first implication of the aforementioned results is that the CR to thermal pressure parameter, which sets the normalization of CRs at the source, may be lower than what is projected for NGC 1068.  Sources with lower values are beyond the reach of the current generation of neutrino telescopes, and their identification would be feasible with the commissioning of IceCube-Gen2 \citep{Kheirandish:2021wkm}.

Both the selection of the X-ray bright Seyfert galaxies in this study and the expected neutrino flux in the disk-corona model, depend strongly on the reported intrinsic X-ray flux in BASS. Therefore, the accuracy of the reported estimates for the intrinsic X-ray emission becomes one of the main hurdles and the primary source of astrophysical systematic uncertainty in this analysis. Among the most promising candidate sources that are considered in this analysis are Compton thick AGN, i.e. AGN with high levels of X-ray obscuration (column density $N_{\rm H} \gtrsim 10^{24} \, \rm cm^{-2} $), for which the assessment of the intrinsic luminosity is challenging. This includes CGCG 420-015 for which the estimated neutrino flux, if interpreted as a genuine signal, would exceed the model expectation by about an order of magnitude. The BASS catalog that is utilized in this analysis offers the most comprehensive survey of non-jetted AGN. However, the accurate measurement of the intrinsic X-ray flux from Compton thick sources requires careful modeling that can benefit from additional data, especially targeting instruments such as {\em NuStar} that are sensitive to higher energy X-rays. It is worth mentioning that detailed modeling of a few of the prominent sources in these searches \citep{Tanimoto:2022wrq} yields an intrinsic flux that is not compatible with BASS. For example, the higher intrinsic flux reported for NGC 1068 \citep{Marinucci:2015fqo} would, compared to BASS, prefer a lower value of CR pressure within the disk-corona model to describe the neutrino flux. Correspondingly, adopting a lower value for the rest of the sources would decrease the expected emission from these other sources in the catalog. Similarly, the discrepancy between different measurements of the intrinsic luminosity for the rest of the sources would change their expected neutrino fluxes. For a recent evaluation of the intrinsic X-ray luminosity of Compton thick AGN, see \citet{Tanimoto:2022wrq}. Additional studies are particularly encouraged, in light of the absence of a detectable neutrino signal in the direction of NGC 4388. Given its proximity, and the high level of intrinsic X-ray flux reported by BASS, the disk-corona model with high CR pressure predicts 21 events from NGC 4388, making it the brightest source in our list excluding NGC 1068. A recent dedicated analysis of the intrinsic emission from this source, however, reports a lower value for the intrinsic flux and a drop during the period of this analysis~\citep{Torres-Alba:2023anq}. A reduction in X-ray emissivity during the IceCube data taking period would lower the predicted neutrino emission and would be consistent with the non-observation of neutrinos reported here for this source \citep{Torres-Alba:2023anq}. In summary, a more accurate measurement of the intrinsic X-ray flux from these sources will likely modify the expected emission from the Seyfert galaxies and the prospects for identifying them.

\section{Summary \& Outlook}
In this work, we present a study of potential high-energy neutrino emission from Seyfert galaxies in the Northern Hemisphere that are intrinsically X-ray bright in X-rays. In addition to the generic power-law flux assumption, we incorporate the disk-corona model and performed a catalog search with $27$ ($28$) sources excluding (including) NGC 1068. We also performed a search for aggregated neutrino emission using a stacking method relying on the disk-corona flux model. As there is no significant excess of neutrino events observed in the stacking search, we can constrain the collective neutrino emission from those X-ray bright Seyfert galaxies in the Northern sky.\\
Since we cannot reject the null hypothesis, we set upper limits on the neutrino emission from all individual sources for both scenarios. The current results motivate continuing searches for neutrino emission from X-ray bright Seyfert galaxies in addition to NGC 1068, especially NGC 4151 and CGCG 420-015, which will reveal whether the cumulative 2.7 sigma excess reported here is due to statistical fluctuations or a genuine astrophysical signal. If interpreted as the latter, it would suggest the existence of sources similar to NGC 1068, that could potentially be explained by the disk-corona model. Nevertheless, the absence of a significant correlation in the stacking search and most individual sources implies that the features of NGC 1068 leading to the strong neutrino emission are not commonly shared with other X-ray bright Seyfert galaxies. As discussed in Sec.~\ref{sec:discussion}, the expectation of neutrino emission relies considerably on the details of the modeling within the picture of the disk-corona model, and more comprehensive multi-wavelength observations will provide further insight into the characteristics of the potential sources which is expected to significantly improve their modeling. 
The results reported here show that implementing dedicated models is useful and can improve the sensitivity of searches for the sources of high-energy neutrinos. 

Our conclusions are supported by a complementary IceCube study of hard X-ray (14-195 keV) AGN, identified in BASS, which reports an excess of neutrinos towards NGC 4151 at 2.9$\sigma$ post-trial significance \citep{IceCube:2024abc}. This analysis is based on an alternative track event selection \citep{IceCube:2019cia} that includes data from the entire sky recorded during partial detector configurations before IceCube was fully commissioned. While this study employs a different hypotheses, data sample, and analysis techniques, the results of the catalog search is consistent with the results reported here.

IceCube-Gen2, the next-generation of the IceCube detector~\citep{IceCube-Gen2:2020qha} will be 8 times larger in volume with an expected 5 times increase of the effective area for muon tracks, increasing the potential to discover neutrino sources by a factor of $\sim 5$. This improvement could lead to the discovery of neutrino emission from the interesting sources studied in this work.

Considering the fact that the majority of bright Seyfert galaxies, the Circinus galaxy for example, reside in the Southern Sky, an enhanced sensitivity towards that region would help the search for more sources similar to NGC 1068. The recent technical progress in starting track and cascade event selections in IceCube ~\citep{IceCube:2021ctg, Abbasi:2024jro, Abbasi:2023bvn} provides significantly improved sensitivity to the Southern Sky, thus creating an excellent opportunity to search for neutrino emission from these interesting Southern Sky sources. In the upcoming years, detectors built, or under construction, in the Northern Hemisphere such as Baikal-GVD, KM3NeT, P-ONE, and TRIDENT~\citep{Baikal-GVD:2018isr,KM3Net:2016zxf,P-ONE:2020ljt,Ye:2022vbk} will further boost the identification of sources in the Southern Sky. 

\section*{Acknowledgements}
The authors gratefully acknowledge the support from the following agencies and institutions:
USA {\textendash} U.S. National Science Foundation-Office of Polar Programs,
U.S. National Science Foundation-Physics Division,
U.S. National Science Foundation-EPSCoR,
U.S. National Science Foundation-Office of Advanced Cyberinfrastructure,
Wisconsin Alumni Research Foundation,
Center for High Throughput Computing (CHTC) at the University of Wisconsin{\textendash}Madison,
Open Science Grid (OSG),
Partnership to Advance Throughput Computing (PATh),
Advanced Cyberinfrastructure Coordination Ecosystem: Services {\&} Support (ACCESS),
Frontera computing project at the Texas Advanced Computing Center,
U.S. Department of Energy-National Energy Research Scientific Computing Center,
Particle astrophysics research computing center at the University of Maryland,
Institute for Cyber-Enabled Research at Michigan State University,
Astroparticle physics computational facility at Marquette University,
NVIDIA Corporation,
and Google Cloud Platform;
Belgium {\textendash} Funds for Scientific Research (FRS-FNRS and FWO),
FWO Odysseus and Big Science programmes,
and Belgian Federal Science Policy Office (Belspo);
Germany {\textendash} Bundesministerium f{\"u}r Bildung und Forschung (BMBF),
Deutsche Forschungsgemeinschaft (DFG),
Helmholtz Alliance for Astroparticle Physics (HAP),
Initiative and Networking Fund of the Helmholtz Association,
Deutsches Elektronen Synchrotron (DESY),
and High Performance Computing cluster of the RWTH Aachen;
Sweden {\textendash} Swedish Research Council,
Swedish Polar Research Secretariat,
Swedish National Infrastructure for Computing (SNIC),
and Knut and Alice Wallenberg Foundation;
European Union {\textendash} EGI Advanced Computing for research;
Australia {\textendash} Australian Research Council;
Canada {\textendash} Natural Sciences and Engineering Research Council of Canada,
Calcul Qu{\'e}bec, Compute Ontario, Canada Foundation for Innovation, WestGrid, and Digital Research Alliance of Canada;
Denmark {\textendash} Villum Fonden, Carlsberg Foundation, and European Commission;
New Zealand {\textendash} Marsden Fund;
Japan {\textendash} Japan Society for Promotion of Science (JSPS)
and Institute for Global Prominent Research (IGPR) of Chiba University;
Korea {\textendash} National Research Foundation of Korea (NRF);
Switzerland {\textendash} Swiss National Science Foundation (SNSF).

\bibliography{bibfile}

\appendix

\begin{figure*}[hpt!]
\centering
\subfigure{%
\includegraphics[width=0.49\textwidth]{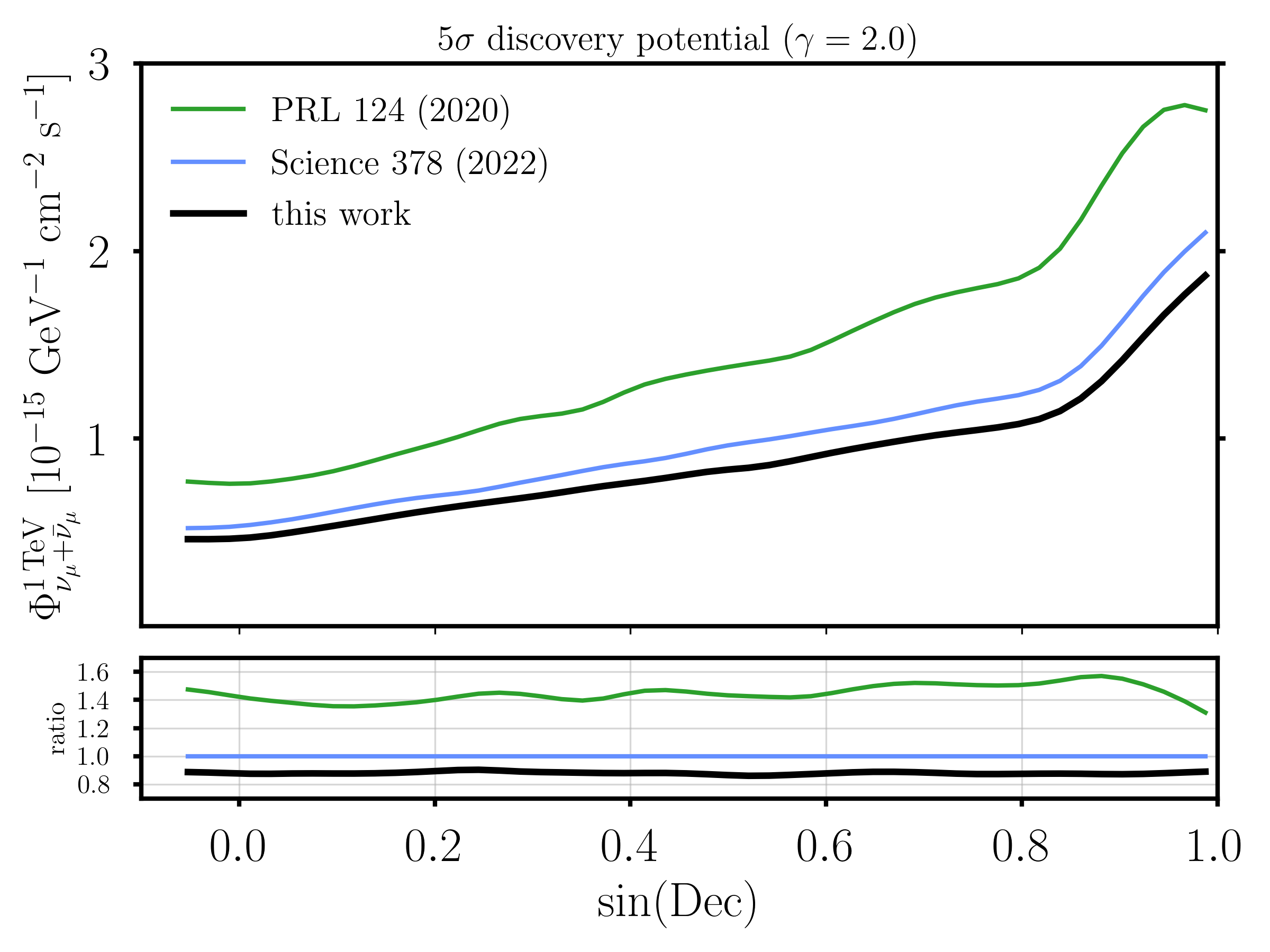}}
\subfigure{%
\includegraphics[width=0.49\textwidth]{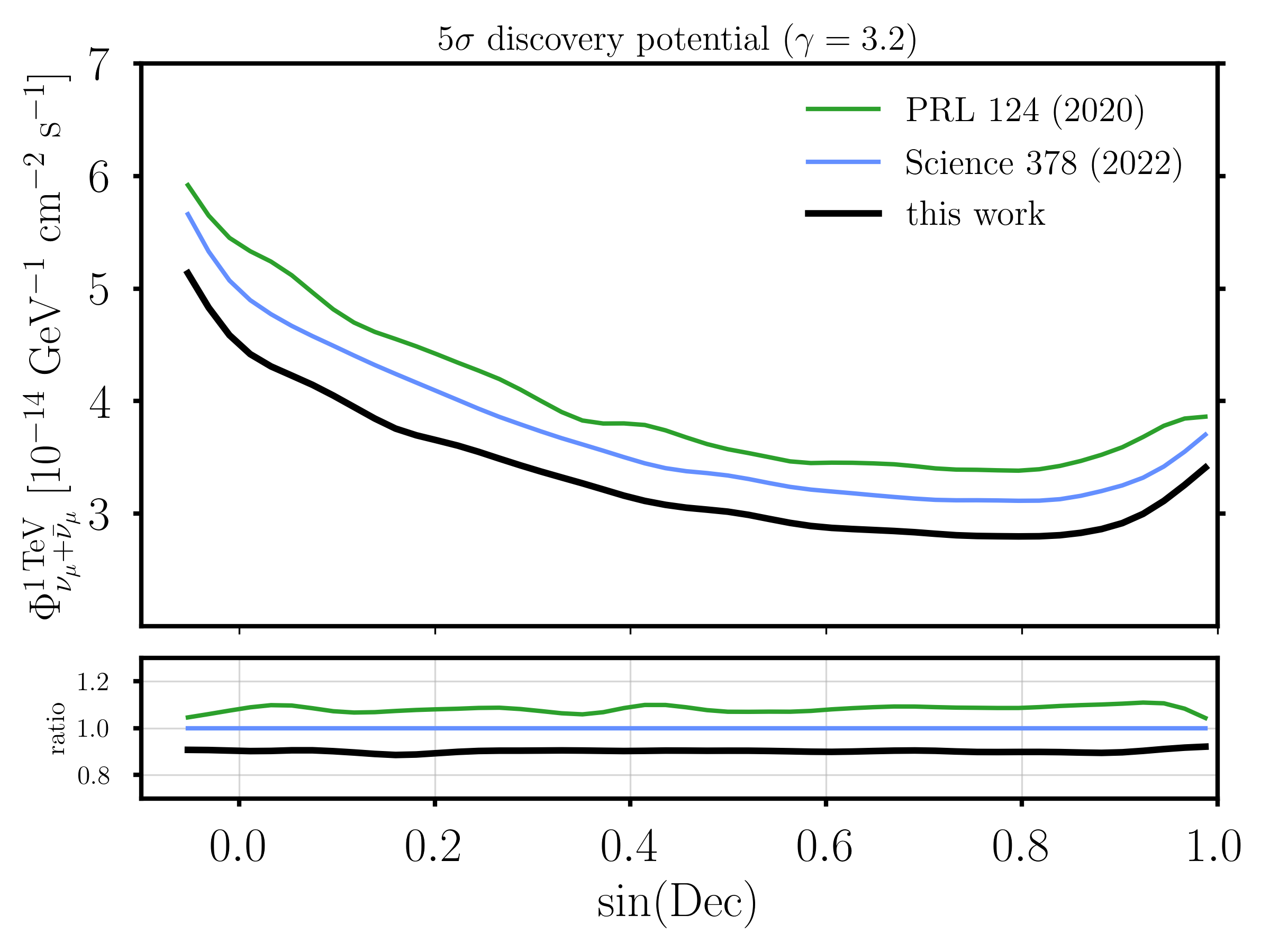}}
\caption{The discovery potential comparison between~\cite{Aartsen:2019fau, IceCube:2022der} and this work for a time-integrated search assuming a power-law spectrum as a function of the source declination. Shown here are the flux levels and the ratios assuming spectral indices $\gamma=2.0$ (left) and $\gamma=3.2$ (right).
}
\label{fig:appendix_fig3}
\end{figure*}

The analysis presented here is an extension of the one described in \citep{IceCube:2022der}. First, we increased the number of neutrino events from $\sim 665,000$ to $\sim 794,000$ by adding data recorded by IceCube during an additional period of $1.7$ years. Second, in addition to the generic power-law spectral assumption, we studied a specific model \citep{Murase:2019vdl, Kheirandish:2021wkm} that relates intrinsic X-ray fluxes to expected neutrino fluxes from the coronal regions of bright Seyfert galaxies. All other technical aspects, such as the event selection criteria and the likelihood function, remain identical. Increasing the available livetime by $\sim 20\%$ improves the $5\sigma$ discovery potential of the analysis by $\sim10\%$, as shown in Fig. \ref{fig:appendix_fig3} for a single power-law spectral assumption with spectral indices $\gamma=2.0$ (left) and $\gamma=3.2$ (right).

\begin{figure*}[t!]
\centering
\subfigure{%
\includegraphics[width=0.32\textwidth]{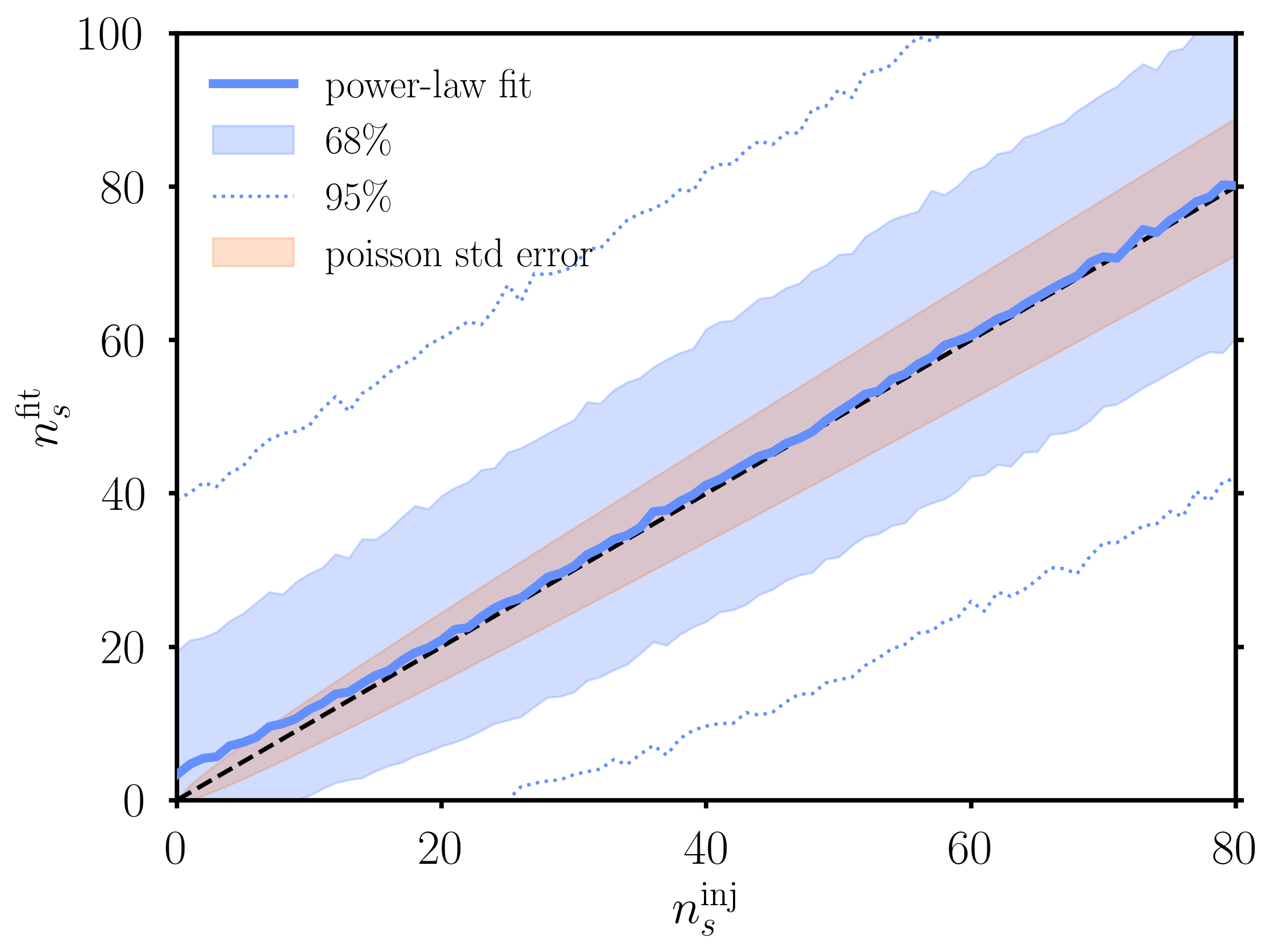}}
\subfigure{%
\includegraphics[width=0.32\textwidth]{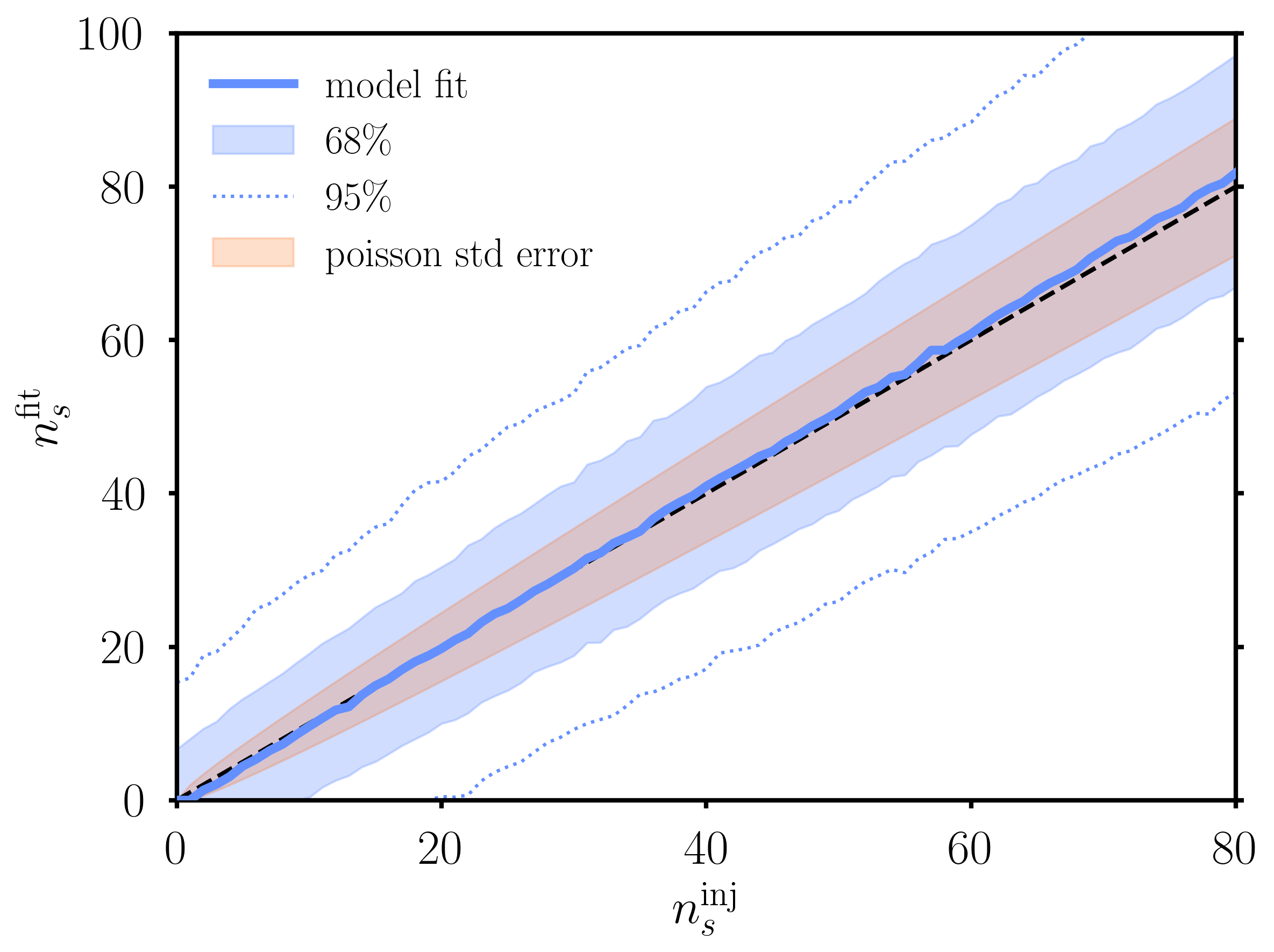}}
\subfigure{%
\includegraphics[width=0.32\textwidth]{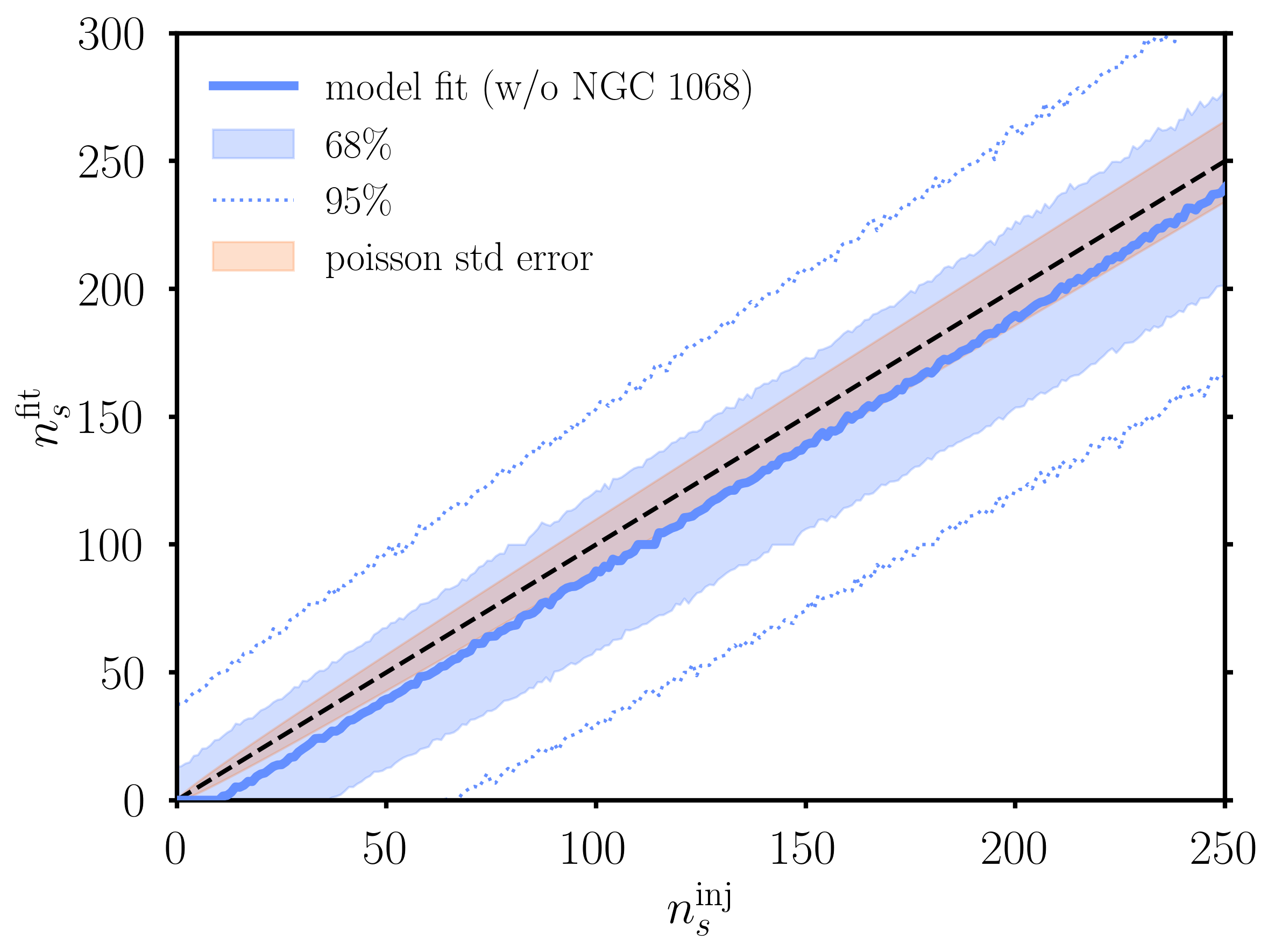}}
\caption{Fitting bias checks. Fitted $n_s$ vs. injected $n_s$ assuming power-law spectrum with $\gamma=3$ (left) and a spectrum predicted by the disk-corona model (middle) for NGC 1068. Same comparison for the stacking search with the disk-corona model flux assumption (right).  
}
\label{fig:appendix_fig4}
\end{figure*}

\begin{figure*}[h!]
\centering
\includegraphics[width=0.5\textwidth]{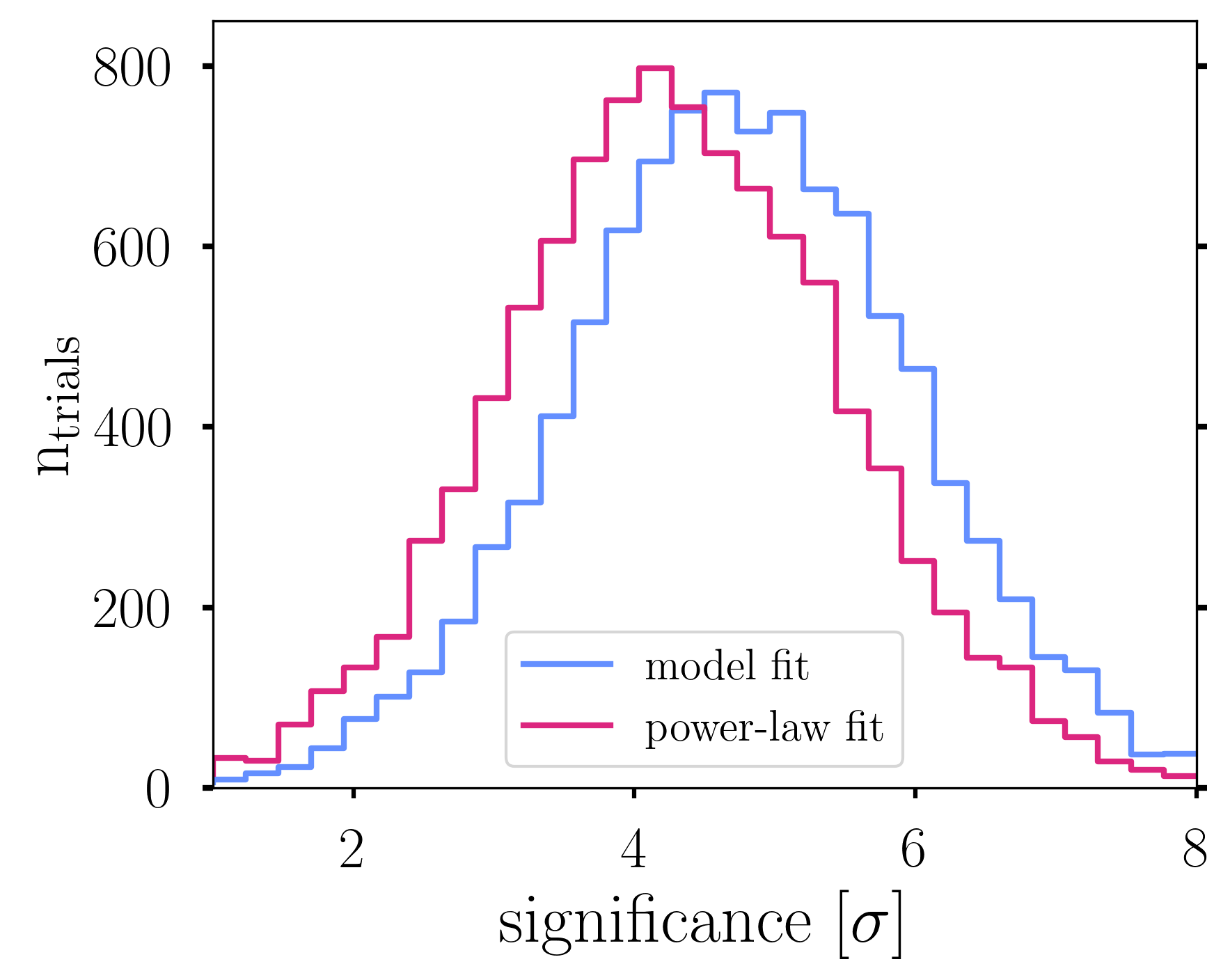}
\caption{The significance comparison between fitting the model (blue) and power-law (orange) spectra for NGC 1068 when simulating the flux predicted by the disk-corona model. The improvement depends on the source and an increase of 0.7$\sigma$ is expected for NGC 1068.}
\label{fig:appendix_fig5}
\end{figure*}

The KDE of the signal and background probability density functions that govern the reconstructed muon directions and energies, developed in \cite{IceCube:2022der}, deliver unbiased parameter estimates for the best-fit number of neutrino events ($n_s$) and power-law spectral indices ($\gamma$), see also Fig. \ref{fig:appendix_fig4} (left). We use the same KDE method to derive the probability density functions assuming the neutrino fluxes predicted by the disk-corona model. Specifically, we construct these KDEs on a discrete grid as a function of the AGN's intrinsic X-ray luminosity. Assuming the intrinsic X-ray luminosities reported by BASS ~\citep{Ricci:2017dhj}, this assigns a unique set of energy and spatial PDFs to each AGN in our selection. Therefore, in contrast to the power-law spectral assumption, there is only a single parameter related to the normalization (the number of signal events $n_s$). The measurement of this parameter remains unbiased, as demonstrated in Fig. \ref{fig:appendix_fig4} (center) for the NGC 1068 as an example. The reduction in variance compared to the power-law case (left) is due to the reduction in degrees of freedom. Assuming the disk-corona model to be correct, we expect the dedicated search to be more powerful than the generic, power-law-based search. 
Simulations show 10-20\% improvement regarding the signal events needed. Fig. \ref{fig:appendix_fig5} shows the increase of the significance using NGC 1068 as an example.
We extend the dedicated, model-based search for neutrinos from individual AGN to the stacking case, i.e. the search for combined signals from the set of Seyfert galaxies selected for this work. The corresponding likelihood function reads

\begin{equation}
    L\left(\mu_s,|\,\boldsymbol{x}\right)=\prod_{i=1}^{N}\left\{\frac{\mu_{s}}{N}\left[\sum_{k=1}^{N_{src}}w_k\,f_s(\boldsymbol{x_i}\,|\,L_k^{\rm 2-10 \, keV},\,\delta_k)\right]+\left(1-\frac{\mu_s}{N}\right)f_b(\boldsymbol{x_i})\right\}
\end{equation}

where the relative weights, $0<w_k<1$, are given by the number of neutrinos expected from IceCube's effective area and the neutrino flux predicted by the disk-corona model for each source depending on the intrinsic X-ray luminosity

\begin{equation}
    w_k=n_k^{exp}(L_k^{\rm 2-10 \, keV}, \delta_k) \times \left\{\sum_k^N{n_k^{exp}\left(L_k^{\rm 2-10 \, keV}, \delta_k\right)}\right\}^{-1}.
\end{equation}

Recovering some fixed number of signal neutrinos from a set of sources is more challenging than recovering the same number of signal neutrinos from a single source, because the effective amount of background is larger. Nevertheless, the estimate of the number of signal events in this analysis remains essentially unbiased also in the stacking case, as shown in Fig.~\ref{fig:appendix_fig4} (right).

To facilitate an easier comparison to previous works, we have also analyzed our selection of Seyfert galaxies assuming a single power-law flux. The measured flux normalizations and spectral indices for the three most-significant sources (NGC 1068, NGC 4151, CGCG 420-015) are given in Fig. \ref{fig:flux_catalog_search}.
The uncertainties presented here are the statistical ones. The detailed study of systematic uncertainties for this sample was presented in the supplemental material of  \citet{IceCube:2022der}. These include the depth-dependent optical properties of the glacial ice, the refrozen hole ice columns around the IceCube strings, as well as the photon detection efficiency of the IceCube DOMs. Their effect on the fitted parameters, studied via dedicated Monte-Carlo simulations of alternative detector properties, is subdominant compared to the statistical uncertainties. For NGC 1068, for which statistical uncertainties are smallest, the systematic uncertainties are at the level of $\sim 10\%$ ($\sim30\%$) of the statistical ones for the mean number of signal events (spectral index). Generally, systematic uncertainties yield $\pm 10\%$ uncertainty on the normalization of the neutrino flux \citep{IceCube:2018ndw}.
The neutrino energy range of the best-fitted flux in Fig. \ref{fig:flux_catalog_search} contains 68\% central bins contribution to the total TS value. We calculate the TS contribution for a given neutrino energy bin by creating a histogram of the weighted sum of neutrino energies selected from similar events in MC (having close angular uncertainty and reconstructed energy values), with weights given by each event's contribution to the total TS value.
The resulting neutrino energy range depends on the best-fitted flux shape. For the power-law flux, we can see how the larger best-fitted $\hat{\gamma}$ spectral index value results in the lower energy range of NGC 1068 in comparison to energy ranges of NGC 4151 and CGCG 420-015, where the best-fitted $\hat{\gamma}$ spectral indices are lower.
The best-fitting values for the entire selection of sources are given in Table \ref{tab:int_results}.\\

\begin{figure*}[hpt!]
\centering
\subfigure{%
\includegraphics[width=0.48\textwidth]{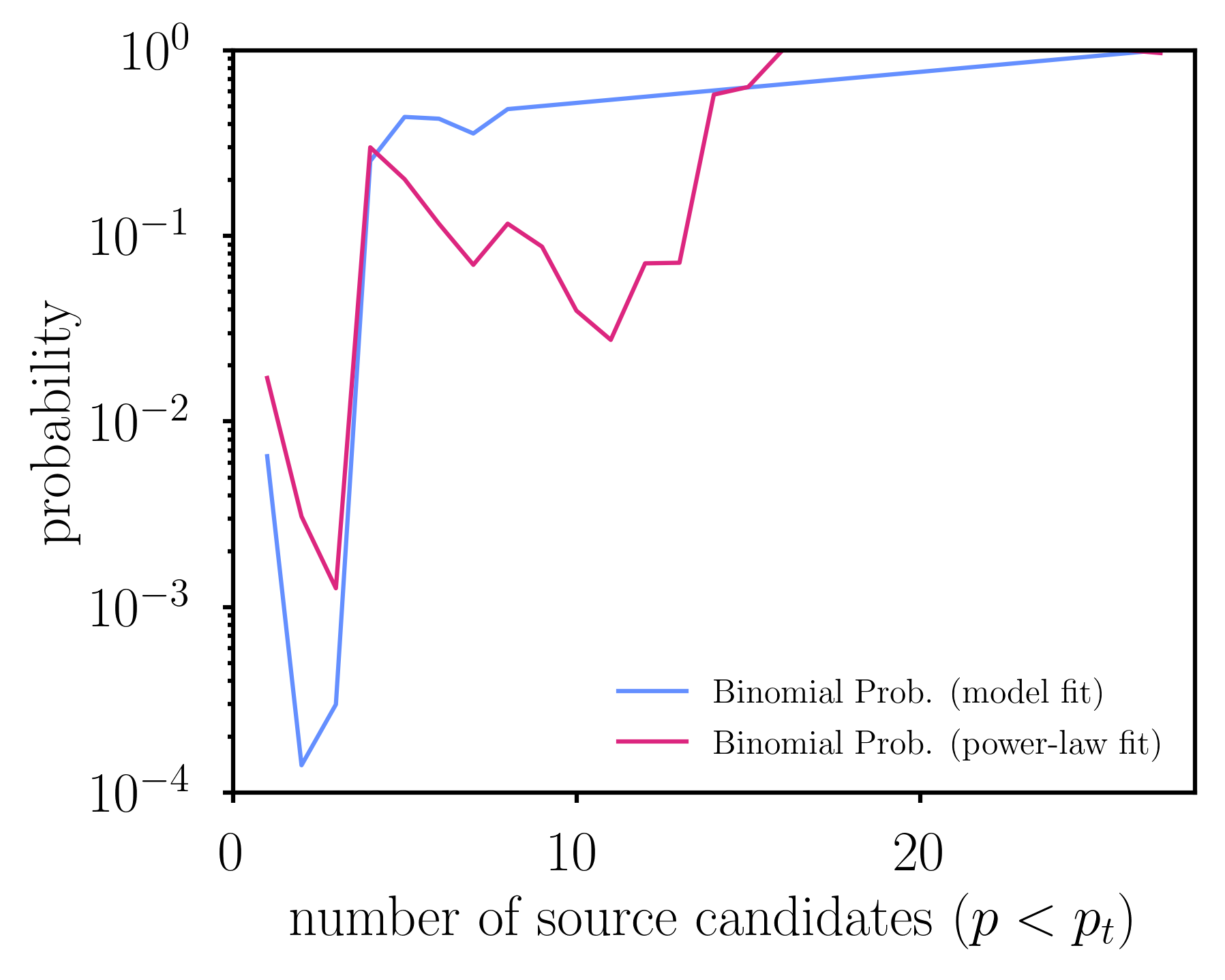}}
\subfigure{%
\includegraphics[width=0.48\textwidth]{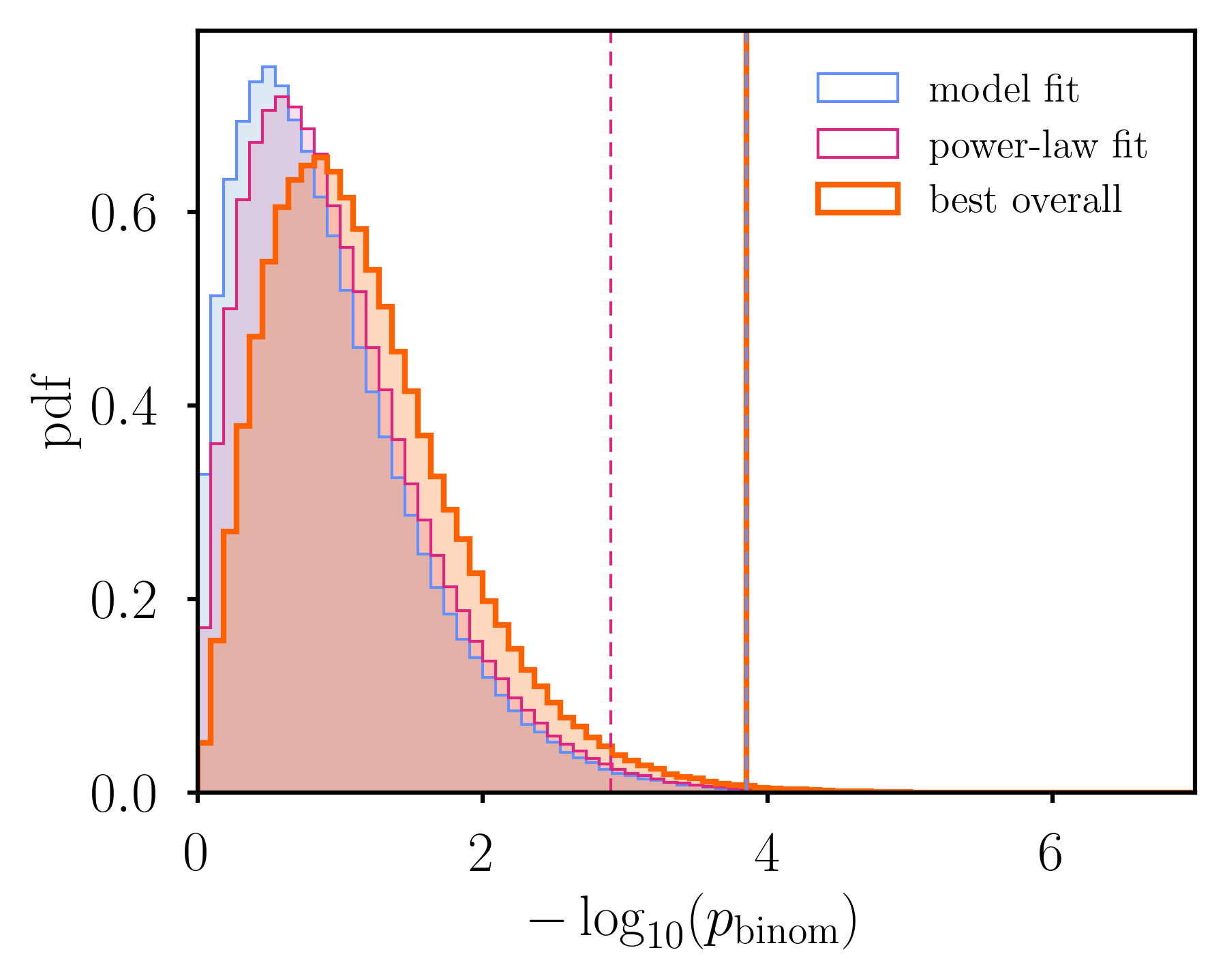}}
\caption{Binomial test for the catalog. {\bf{Left}}: Scans of the $p$-value threshold for the disk-corona model (red) and the power-law spectrum (blue). The strongest excess is found at $k=2$ (NGC 4151 and CGCG 420-015) assuming the disk-corona model flux. {\bf{Right}}: Distributions of binomial $p$-value in background simulations. The vertical dashed lines denote the observed binomial $p$-values with the experimental data assuming the power-law flux (red) and disk-corona model flux (blue). The trial-corrected distribution corresponds to the orange histogram and the final $p$-value is denoted by the orange line.
}
\label{fig:binomial}
\end{figure*}

A search for aggregated neutrino signals using the stacking method described above can fail if the model misspecifies individual candidate sources, but is otherwise mostly correct.
The binomial test trades a reduction in sensitivity for increased robustness. It simply considers a subset of sources that show positive results in the single source analysis.
Here, we perform it  twice. One binomial test for each of the two spectral assumptions: disk-corona model and single power-law. The test uses the pre-trial $p$-values of the sources to examine an excess in the number of small $p$-values compared to the expectation from the background. 
The probability of producing $k$ or more sources with $p$-values smaller than $p_k$ from background is 
\begin{equation}
p_{bkg} = \sum_{i=k}^{N_{src}}\binom{N_{\rm src}}{i}p_k\left(1-p_k\right)^{N_{src}-i},
\end{equation}
where we search for the minimum probability $p_{min}$ and its corresponding $k$. The expected distribution of $p_{min}$ for the background used to evaluate the significance is estimated from pseudo experiments. Fig. \ref{fig:binomial} (left) shows the binomial probability as a function of the number of sources that exceed a given local $p$-value threshold for the power-law flux (blue) and the disk-corona model flux (red) assumptions. Overall, the smallest binomial probability $p_{min}=1.4\times10^{-4}$ is found for $k=2$ sources (NGC 4151 and CGCG 420-015) assuming the disk-corona model flux. This does not determine the final significance, as we have to account for the internal trials related to the scan over the local $p$-value thresholds, as well as the choice of two flux models. To determine the final $p$-value, we perform a large number of repetitions of this search using simulated data that containing only background events. Fig. \ref{fig:binomial} (right) shows the distribution of the best binomial $p$-value under the power-law spectral and the corona-disk model assumptions.
Having scanned the threshold local $p$-value during the binomial scan increases the $p$-value from $p_{min}=1.4\times10^{-4}$ to $p_{min}^{corr}=1.7\times10^{-3}$. Selecting the best solution among the two flux assumptions further increases the $p$-value to $p_{final}=3.4\times10^{-3}$, or $2.7\sigma$ (orange).

\begin{figure*}[t!]
\centering
\subfigure{%
\includegraphics[width=0.32\textwidth]{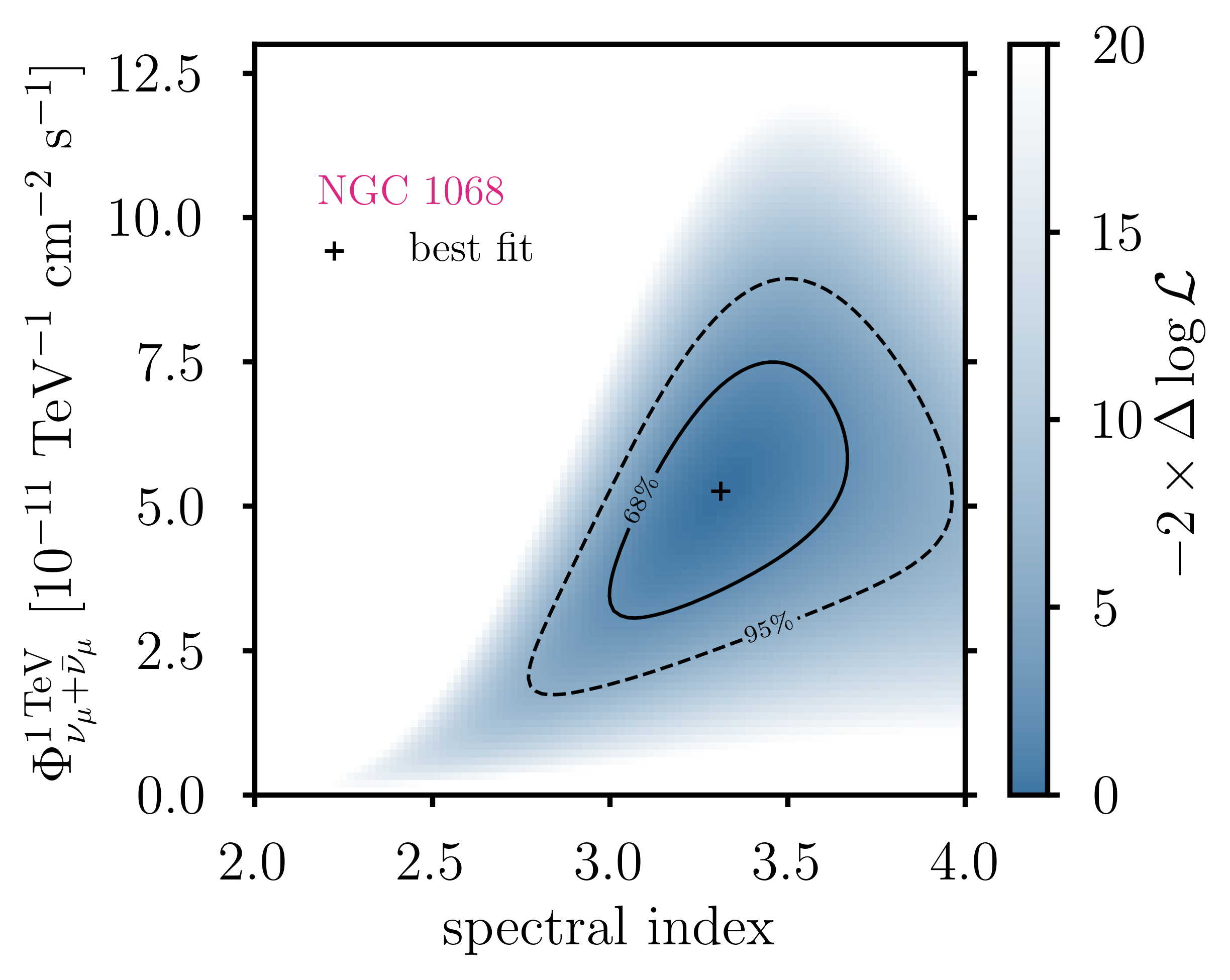}}
\subfigure{%
\includegraphics[width=0.32\textwidth]{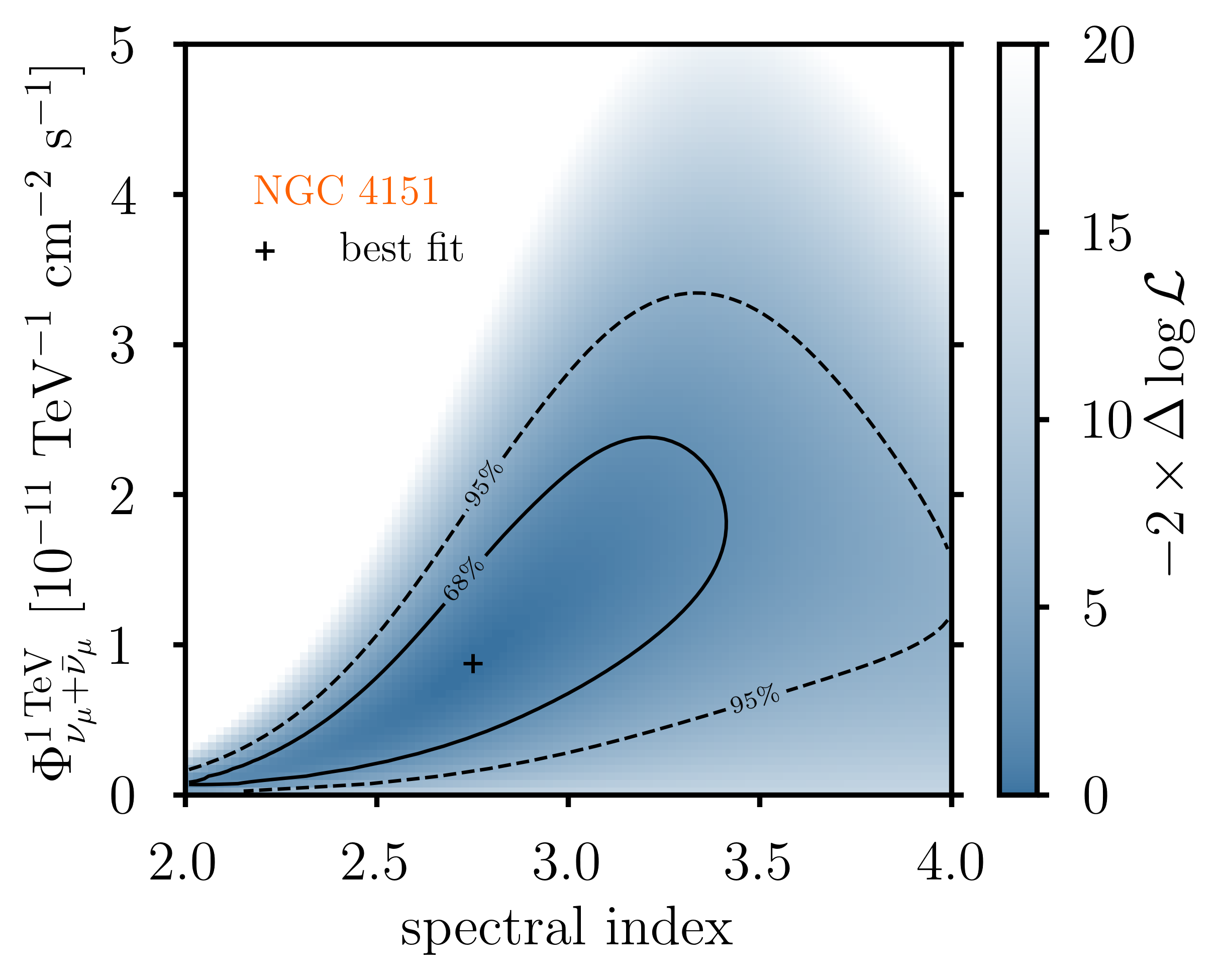}}
\subfigure{%
\includegraphics[width=0.32\textwidth]{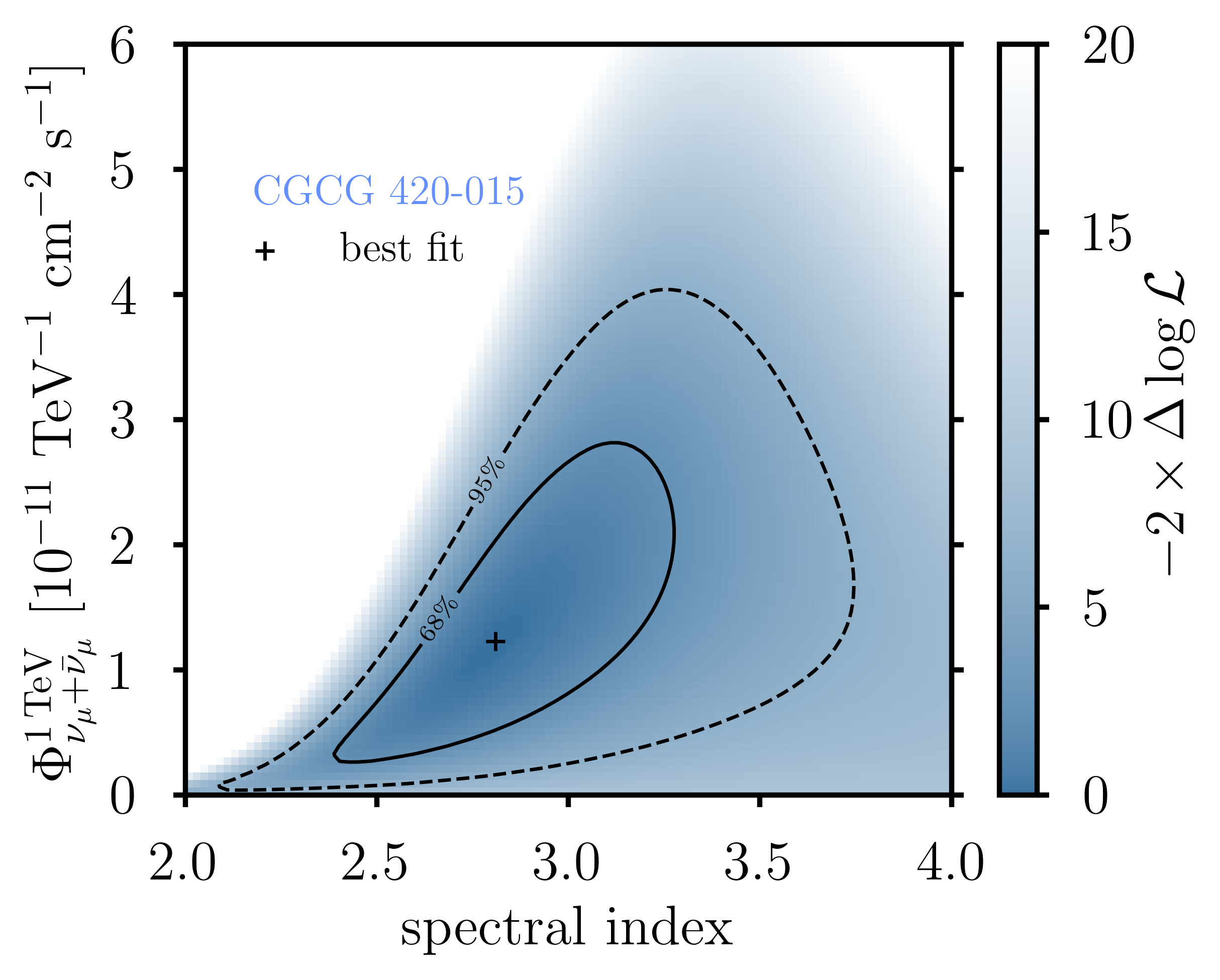}}
\caption{Profile likelihood scans for the flux parameters for the most significant sources, assuming a power-law spectrum: NGC 1068 (left), NGC 4151 (middle), and CGCG 420-015 (right). The crosses show the best-fit values while contours correspond to the 68\% (solid) and 95\% (dashed) confidence regions. }
\label{fig:spectrum_scan}
\end{figure*}

\begin{deluxetable}{lDDD|DDDD|DDDDDD}[h]
\tablecaption{Source List and Results}
\tablewidth{9pt}
\tablehead{
\colhead{}  & \multicolumn2c{} &  \multicolumn2c{} & \multicolumn2c{}  & \multicolumn2c{} & \multicolumn2c{model}&\multicolumn2c{}&\multicolumn2c{} 
&\multicolumn2c{}&\multicolumn2c{}&\multicolumn2c{power-law}&\multicolumn2c{}&\multicolumn2c{}\\
\colhead{Source}  & \multicolumn2c{Decl.} &  \multicolumn2c{R.A.} & \multicolumn2c{$F_{2-10\mathrm{keV}}^{\rm intr}$}  & \multicolumn2c{$n_{\rm exp}$} & \multicolumn2c{$\hat{n}_{\rm s}$}&\multicolumn2c{-log$_{10} p$}&\multicolumn2c{$n_{\rm UL}$} 
&\multicolumn2c{$\hat{n}_s$}&\multicolumn2c{$\hat{\gamma}$}&\multicolumn2c{-log$_{10}p$}&\multicolumn2c{$\phi^{E^{-2}}_{90\%}$} &\multicolumn2c{$\phi^{E^{-3}}_{90\%}$}
}  
\decimals
\startdata
NGC 1068 & -0.0 & 40.7 & 268.3 & 44.5 & 47.5 & 6.5 & 61.4  & 94.1  & 3.3 & 7.1 & 8.5 & 39.0 \\
NGC 4388 & 12.7 & 186.4 & 71.7 & 21.4 & 0.0 & 0.0 & 13.0  & 2.0  & 1.9 & 0.9 & 3.9 & 16.7 \\
NGC 6240 & 2.4 & 253.2 & 411.1 & 16.8 & 0.0 & 0.0 & 13.4  & 0.0  & 4.3 & 0.0 & 1.5 & 5.8 \\
NGC 4151 & 39.4 & 182.6 & 84.8 & 13.1 & 22.5 & 3.2 & 39.5  & 30.1  & 2.7 & 3.2 & 10.9 & 38.7 \\
Z164-19 & 27.0 & 221.4 & 179.5 & 8.6 & 0.0 & 0.0 & 12.0  & 3.3  & 2.0 & 0.7 & 4.2 & 15.7 \\
UGC 11910 & 10.2 & 331.8 & 157.5 & 8.5 & 0.0 & 0.0 & 12.9  & 6.4  & 4.3 & 0.3 & 2.2 & 8.5 \\
NGC 5506 & -3.2 & 213.3 & 115.6 & 8.1 & 0.0 & 0.0 & 9.0  & 0.0  & 1.6 & 0.0 & 1.9 & 6.4 \\
NGC 1194 & -1.1 & 46.0 & 117.8 & 7.6 & 4.4 & 0.6 & 15.2  & 27.7  & 3.7 & 0.9 & 2.9 & 13.1 \\
Mrk3 & 71.0 & 93.9 & 113.6 & 7.4 & 0.0 & 0.0 & 10.9  & 0.0  & 4.3 & 0.0 & 4.4 & 11.4 \\
MCG+8-3-18 & 50.1 & 20.6 & 99.4 & 6.3 & 0.0 & 0.0 & 10.8  & 0.0  & 4.3 & 0.0 & 3.3 & 9.3 \\
UGC 3374 & 46.4 & 88.7 & 65.1 & 4.6 & 0.0 & 0.0 & 11.0  & 0.0  & 4.3 & 0.0 & 3.2 & 9.0 \\
NGC 3227 & 19.9 & 155.9 & 37.2 & 4.0 & 0.0 & 0.0 & 14.5  & 0.0  & 1.7 & 0.0 & 2.1 & 6.8 \\
4C+50.55 & 51.0 & 321.2 & 97.0 & 4.0 & 4.6 & 0.8 & 14.9  & 9.7  & 3.2 & 0.5 & 5.0 & 15.9 \\
NGC 7682 & 3.5 & 352.3 & 47.9 & 4.0 & 2.3 & 0.7 & 18.8  & 0.0  & 4.3 & 0.0 & 1.6 & 6.2 \\
IRAS05078+1626 & 16.5 & 77.7 & 46.1 & 4.0 & 0.0 & 0.0 & 12.2  & 0.0  & 4.3 & 0.0 & 2.0 & 6.9 \\
2MASXJ20145928+2523010 & 25.4 & 303.7 & 78.6 & 3.8 & 0.0 & 0.0 & 11.9  & 0.0  & 4.3 & 0.0 & 2.3 & 7.6 \\
Mrk 1040 & 31.3 & 37.1 & 40.6 & 3.7 & 0.0 & 0.0 & 11.7  & 32.9  & 4.3 & 0.9 & 5.1 & 19.1 \\
LEDA136991 & 68.4 & 6.4 & 42.6 & 3.7 & 0.0 & 0.0 & 11.4  & 3.8  & 4.1 & 0.2 & 5.0 & 13.4 \\
Mrk 1210 & 5.1 & 121.0 & 32.9 & 3.2 & 0.0 & 0.0 & 13.3  & 0.0  & 4.3 & 0.0 & 1.7 & 6.4 \\
CGCG 420-015 & 4.1 & 73.4 & 50.5 & 3.2 & 30.7 & 3.6 & 46.4  & 35.5  & 2.8 & 2.5 & 5.2 & 25.9 \\
MCG+4-48-2 & 25.7 & 307.1 & 31.6 & 3.1 & 22.1 & 2.3 & 31.8  & 45.2  & 3.2 & 2.1 & 7.2 & 29.0 \\
3C111 & 38.0 & 64.6 & 61.5 & 3.1 & 0.0 & 0.0 & 11.6  & 15.7  & 4.3 & 0.5 & 4.2 & 13.6 \\
UGC 5101 & 61.4 & 144.0 & 45.4 & 2.6 & 4.8 & 1.0 & 17.6  & 8.7  & 3.0 & 0.7 & 6.9 & 21.7 \\
3C382 & 32.7 & 278.8 & 49.4 & 2.4 & 0.0 & 0.0 & 11.6  & 34.9  & 4.3 & 1.0 & 5.4 & 20.1 \\
Mrk 110 & 52.3 & 141.3 & 34.4 & 2.1 & 0.0 & 0.0 & 10.9  & 0.0  & 4.3 & 0.0 & 3.4 & 9.6 \\
3C 390.3 & 79.8 & 280.5 & 44.4 & 1.8 & 0.0 & 0.0 & 12.6  & 0.0  & 4.3 & 0.0 & 6.9 & 19.7 \\
NGC 3516 & 72.6 & 166.7 & 30.7 & 1.6 & 0.0 & 0.0 & 11.8  & 30.0  & 4.3 & 0.6 & 8.8 & 26.0 \\
Cygnus A & 40.7 & 299.9 & 32.1 & 1.6 & 3.7 & 0.7 & 15.2  & 2.9  & 2.1 & 0.7 & 5.3 & 18.2 \\
\enddata
\tablecomments{Information of sources and the catalog search results. Intrinsic 2-10~keV X-ray flux is $F_{2-10\mathrm{keV}}^{\rm intr}\,\times 10^{-12}\,\mathrm{erg\,cm^{-2}s^{-1}}$. Best-fit results for TS, $\hat{n}_s$ and pre-trial $p$-values  for both the model analysis and power-law spectral assumptions are shown. For the model analysis, expected numbers of signal events ($n_{\rm exp}$) and 90$\%$ upper limit of the signal event numbers ($n_{\rm UL}$) are listed and for the power-law analysis, best-fitted spectral indices $\hat{\gamma}$ and 90$\%$ upper limit fluxes are listed. The upper limit fluxes are parameterized as $\phi^{E^{-\gamma}}_{90\%}  \left(E/ 1\,\mathrm{TeV}\right)^{-\gamma}\times 10^{-13}\, {\mathrm{TeV^{-1}cm^{-2}s^{-1}}}$. 
} 
\label{tab:int_results}
\end{deluxetable}

\end{document}